\documentclass[prd,showpacs,showkeys,nofootinbib,twocolumn]{revtex4-1}
\usepackage{amsmath,amsfonts,amssymb,color}
\usepackage{bm}
\usepackage{pgfplots}
\usepackage[colorlinks=true,urlcolor=blue,linkcolor=blue,citecolor=blue]{hyperref}
\usepackage{color}
\newcommand{\aQ}{\nearrow\!\!\!\!\!\!\!Q}
\newcommand{\aZ}{\nearrow\!\!\!\!\!\!\!Z}
\newcommand{\pW}{\stackrel {(+)}W}
\newcommand{\pmW}{\stackrel {(\pm)}W}
\newcommand{\pOm}{\stackrel {(+)}\Omega}
\newcommand{\mOm}{\stackrel {(-)}\Omega}
\newcommand{\pmOm}{\stackrel {(\pm)}\Omega}

\begin{document} 

\title{Gravitational waves in metric-affine gravity theory}

\author{Alejandro Jim\'enez-Cano}
\email{alejandrojc@ugr.es}
\affiliation{Departamento de F\'{\i}sica Te\'orica y del Cosmos and CAFPE
Universidad de Granada, 18071, Granada, Spain}

\author{Yuri N. Obukhov}
\email{obukhov@ibrae.ac.ru}
\affiliation{Theoretical Physics Laboratory, Nuclear Safety Institute, 
Russian Academy of Sciences, B.Tulskaya 52, 115191 Moscow, Russia}

\begin {abstract}
We derive the exact gravitational wave solutions in a general class of quadratic metric-affine gauge gravity models. The Lagrangian includes all possible linear and quadratic invariants constructed from the torsion, nonmetricity and the curvature. The ansatz for the gravitational wave configuration and the properties of the wave solutions are patterned following the corresponding ansatz and the properties of the plane-fronted electromagnetic wave. 
\end{abstract}


\maketitle

\section{Introduction}

In contrast to Newton's gravity with its absolute space and time, the modern understanding of relativistic gravitational phenomena is based on the idea that the structure and the dynamics of the geometry of the spacetime continuum is determined by the physical matter. As Einstein wrote in \cite{Ein}:   ``...The question whether this continuum has a Euclidean, Riemannian, or any other structure is a question of physics proper which must be answered by experience, and not a question of a convention to be chosen on grounds of mere expediency.'' Classical experimental tests in terrestrial laboratories and observations in the solar system demonstrate the validity of Einstein's general relativity (GR) theory on the macroscopic scales when the matter is characterised by its mass and energy. 

Such a satisfactory status of GR as a macroscopic theory of gravity, however, is not a guarantee that it also correctly describes the gravitational phenomena at extremely small scales when one takes into account that matter is characterized not only by the energy-momentum current but also by other intrinsic properties known as microstructure (spin, shear and dilaton current, in particular). In this respect, an alternative viable description of the gravitational interaction in the microworld (and at earliest stages of universe's evolution) is provided by the gauge approach to gravity \cite{MAG,Blag,reader,PBO}. The gauge principle is one of the cornerstones of the modern physics, which explains the nature of all physical interactions in a consistent field-theoretic Yang-Mills-Higgs framework that is solidly substantiated by high-energy experiments.

For gravity, the corresponding gauge-theoretic formalism can be developed \cite{Sciama,Kibble} along the same lines as for the electroweak and strong interactions by replacing the underlying non-Abelian group of {\it internal} symmetries with a group of {\it spacetime} symmetries, e.g., translational, Lorentz, Poincar\'e, conformal, general linear, or affine one. In particular, the metric-affine gravity (MAG) arises as a gauge theory based on the general affine group $GA(4,R) = T_4\,\rtimes\,GL(4,R)$, a semidirect product of the translation group $T_4$ times the general linear group $GL(4,R)$, when the matter is characterized by the three Noether currents: the canonical energy-momentum current, the canonical hypermomentum current, and the metric energy-momentum current \cite{MAG}. These matter sources are minimally coupled to the corresponding gravitational field potentials: the coframe, the linear connection and the metric, respectively. It is worthwhile to mention that Einstein's GR can be consistently interpreted as a gauge theory under the assumption of a nonminimal coupling of a certain form \cite{grpg}. 

In MAG, the geometrical structure of spacetime is extended from the Riemannian geometry of Einstein's GR to include nontrivial post-Riemannian structures such as the torsion and the nonmetricity. The resulting metric-affine geometry is of interest, both mathematically and physically, for a number of reasons \cite{Goenner,Coley1,Coley2,Coley3,mccrea:1992,Vitagliano}. A strong motivation comes from the geometrical approach to the physics of hadrons in terms of extended structures \cite{nee1,nee2,nee3,MAG}, and from the efforts to construct a consistent quantum gravity theory \cite{lee1,lee2,per1,per2}. The theory of continuous media with microstructure \cite{frank} gives rise to a realistic model of classical matter with hypermomentum \cite{hyper} which is widely used for the study of the early universe's evolution, also relating the post-Riemannian structures to the dark matter problem \cite{dirk3,dirk4,dirk5}. It is worthwhile to mention that certain special MAG models may arise as the effective theories in the dilaton-axion-metric low-energy limit of the string theory \cite{dil1,dil2,dil3,dil4}. It is important to notice that it is possible to detect the post-Riemannian spacetime geometry only with the help of the matter with microstructure \cite{nee,yass,Puetz,eom2}.

The study of the exact solutions of the MAG field equations is important for understanding and development of the physical aspects mentioned above, in order to fix the structure of the basic Lagrangian of the theory, as well as for the detailed analysis of possible new physical effects. The derivation of new exact solutions for these models would bring new insight to the understanding of gravitational physics on microscopic scales, under an important condition of consistency with Einstein's GR at large distances which should be recovered in a certain limit \cite{pono,Gronwald:1997,hehl:1999}. The earlier results include the construction of the spherically and axially symmetric solutions, including the black hole configurations which can carry nontrivial shear and dilaton charges, in addition to the mass \cite{tres1,tres2,wang,vlach:1996,obukhov:1996,Garcia:1998,obukhov:1997,Delhom:2019}. Among other methods, the so-called triplet ansatz technique has proven to be an effective method of deriving exact solutions in MAG \cite{obukhov:1997}.

Wave is a fundamental physical phenomenon, and the gravitational wave research became a rapidly developing subject after the recent experimental discovery of the first gravitational wave signals \cite{Abbott1,Abbott2}. The plane-fronted gravitational waves represent an important class of exact solutions which generalize the basic properties of electromagnetic waves in flat spacetime to the case of curved spacetime geometry. In the framework of GR, the theoretical study of the gravitational waves has a long and rich history \cite{flan,schutz,CNN,Brink1,Brink2,Brink3,rosen1937,einrosen,rosen1956,rosen1958,Virb1,Virb2,bondi0,bondi1,peres,pen1,pen2,Kom1,Kom2,Jordan1,Jordan2,kundt,curr,schim,AT,piran,MashQ,torre,cropp1,cropp2,coley12,mcnutt,Barnett,griff,vdz,exact}. A wide variety of exact gravitational plane wave solutions was obtained in the Poincar\'e gauge gravity \cite{adam,chen,sippel,vadim,singh,babu,BC1,BC2,BC3,BC4,BC5,yno:2017,BCO}, in teleparallel gravity \cite{tele,Conroy:2018,Hohmann:2018,Hohmann:2019,Capozziello:2020,Cai:2016}, in a number of modified gravity theories \cite{gurses,lovelock,Baykal,Mohseni}, as well as in supergravity \cite{sg1,sg2,sg3,sg4,sg5} and in superstring theories \cite{gimon,ark1,ark2,str1,str2,str3,str4}. The higher-dimensional generalizations of the gravitational wave solutions were discussed in \cite{sokol,coley1,coley2,hervik,ndim}.

A critical analysis of the gravitational wave criteria \cite{vdz} was performed in the recent work \cite{AJC}, and an appropriate extension was proposed for the metric-affine spacetimes. The earlier studies \cite{ppmag,dirk1,dirk2,king,vas1,vas2,vas3,pasic1,pasic2} had demonstrated the existence of the gravitational wave solutions in the metric-affine theory of gravity with the propagating torsion and nonmetricity fields. In many cases, however,  either the torsion waves were revealed, or the wave field configurations were found for a special class of the MAG Lagrangian by means of the triplet technique \cite{obukhov:1997,hehl:1999} with a specific ansatz for torsion and nonmetricity. 

The aim of this paper is to describe the plane gravitational waves for the general Yang-Mills type quadratic MAG Lagrangian with nontrivial torsion and nonmetricity configurations that do not belong to the triplet ansatz. The motivations are as follows. Quite generally, the systematic study of the space of solutions represents a significant aspect of the development of any field-theoretic model. At the same time, since the wave phenomena are of fundamental importance as such, the construction and comparison of the wave solutions in different models may help to establish their physical contents and clarify the relations between the microscopic and macroscopic gravitational theories (in particular, general relativity, Poincar\'e gauge gravity and MAG). Moreover, the analysis of the plane wave solutions can provide a good understanding of the particle spectrum for the general quadratic MAG models, extending the earlier results \cite{Karananas,BC6,Baikov:1992,Percacci:2020,Lasenby1,Lasenby2,Lasenby3}. 

The structure of the paper is as follows. In the next Sec.~\ref{MAG}, we give a short overview of the general structure of the MAG theory. Then in Sec.~\ref{GW} we formulate the corresponding ansatz for a gravitational plane wave in MAG. The properties of the resulting curvature, torsion and nonmetricity in terms of their irreducible parts are discussed in Sec.~\ref{irreps}, and the explicit field equations are derived in Sec.~\ref{MAGE} for the general quadratic MAG model (\ref{lagrV}). Finally, in Sec.~\ref{FE} we derive the exact solutions of the MAG field equations for the proposed ansatz. The conclusions are outlined in Sec~\ref{DC}.

\subsection{Notations}

Our basic notation and conventions are consistent with \cite{MAG}. In particular, Greek indices $\alpha, \beta, \dots = 0, \dots, 3$, denote the anholonomic components (for example, of a coframe $\vartheta^\alpha$), while the Latin indices $i,j,\dots =0,\dots, 3$, label the holonomic components ($dx^i$, e.g.). The anholonomic vector frame basis $e_\alpha$ is dual to the coframe basis in the sense that $e_\alpha\rfloor\vartheta^\beta = \delta_\alpha^\beta$, where $\rfloor$ denotes the interior product. The volume 4-form is denoted $\eta$, and the $\eta$-basis in the space of exterior forms is constructed with the help of the interior products as $\eta_{\alpha_1 \dots\alpha_p}:= e_{\alpha_p}\rfloor\dots e_{\alpha_1}\rfloor\eta$, $p=1,\dots,4$. They are related to the $\theta$-basis via the Hodge dual operator $^*$, for example, $\eta_{\alpha\beta} = {}^*\!\left(\vartheta_\alpha\wedge\vartheta_\beta\right)$. We will mark the parity-odd variables by the overline in order to distinguish them from the parity-even objects (for example, $\overline{T}$ denotes the axial trace 1-form of the torsion, whereas $T$ is the torsion trace 1-form).

\section{MAG: brief overview}\label{MAG}

The metric-affine gravity (MAG) is constructed as the gauge theory for the general affine spacetime symmetry group \cite{MAG}. The gravitational field potentials are the metric $g_{\alpha\beta}$, the coframe $\vartheta^\alpha = e^\alpha_i dx^a$ and connection $\Gamma_\alpha{}^\beta = \Gamma_{i\alpha}{}^\beta dx^i$ 1-forms. The corresponding gauge field strengths are identified with the nonmetricity 1-form, the torsion 2-form, and the curvature 2-form, respectively:
\begin{eqnarray}
Q_{\alpha\beta} &=& -\,Dg_{\alpha\beta} = -\,dg_{\alpha\beta} + 2\Gamma_{(\alpha\beta)},\label{nonm}\\
T^\alpha &=& D\vartheta^\alpha = d\vartheta^\alpha +\Gamma_\beta{}^\alpha\wedge
\vartheta^\beta,\label{Tor}\\ \label{Cur}
R_\alpha{}^\beta &=& d\Gamma_\alpha{}^\beta + \Gamma_\gamma{}^\beta\wedge\Gamma_\alpha{}^\gamma.
\end{eqnarray}
As usual, the covariant differential is denoted $D$.

The gravitational Lagrangian 4-form 
\begin{equation}
V = V(g_{\alpha\beta}, \vartheta^{\alpha}, Q_{\alpha\beta}, T^{\alpha}, R_{\alpha}{}^{\beta})\label{lagrV}
\end{equation}
is an arbitrary function of the gravitational field variables. The MAG field equations are derived from the variational derivatives with respect to the coframe and connection. They read explicitly (with the canonical energy-momentum $\Sigma_\alpha$ and the hypermomentum $\Delta^\alpha{}_\beta$ currents as the matter sources):
\begin{eqnarray}
{\frac{\delta V}{\delta\vartheta^{\alpha}}} = 
- DH_{\alpha}  + E_{\alpha} &=& \Sigma_\alpha, \label{dVt}\\ 
{\frac{\delta V}{\delta\Gamma_\alpha{}^\beta}} 
= - DH^\alpha{}_\beta + E^\alpha{}_\beta &=& \Delta^\alpha{}_\beta.\label{dVG}
\end{eqnarray}
These are the 1st, and the 2nd MAG field equations \cite{MAG}. We do not write down the 0th field equation (which arises from the variation with respect to the metric) because it is identically satisfied in view of (\ref{dVt}) and (\ref{dVG}) and the Noether identities (see Sec. 5.5 in \cite{MAG}). Here the partial derivatives of the Lagrangian with respect to the generalized ``velocities''
\begin{eqnarray}
M^{\alpha\beta} &=& -\,2{\frac {\partial V}{\partial Q_{\alpha\beta}}},\label{Mab}\\
H_{\alpha} &=& -\,{\frac {\partial V}{\partial T^{\alpha}}},\label{Ha}\\
H^{\alpha}{}_{\beta} &=& -\,{\frac {\partial V}{\partial R_{\alpha}{}^{\beta}}},\label{Hab}
\end{eqnarray}
are identified as the gravitational field momenta, and
\begin{eqnarray}
E_{\alpha} &=& e_{\alpha}\rfloor V + (e_{\alpha}\rfloor T^{\beta})\wedge H_{\beta}
+ (e_{\alpha}\rfloor R_{\beta}{}^{\gamma})\wedge H^{\beta}{}_{\gamma}\nonumber\\
&& + \,{\frac 12}(e_{\alpha}\rfloor Q_{\beta\gamma})M^{\beta\gamma},\label{Ea}\\
E^{\alpha}{}_{\beta} &=& - \,\vartheta^{\alpha}\wedge H_{\beta} - M^{\alpha}{}_{\beta}.\label{Eab}
\end{eqnarray}
are the canonical gauge field currents of the gravitational energy-momentum and hypermomentum, respectively. 

\subsection{Quadratic metric-affine gravity models}

The 1-form of nonmetricity can be decomposed into 4 irreducible parts, the torsion 2-form can be decomposed into the 3 irreducible parts, whereas the curvature 2-form has 11 irreducible pieces. Their definition is presented in Appendices~\ref{irrtor}-\ref{irrcur}.

The general quadratic model is described by the Lagrangian 4-form that contains all possible quadratic invariants of the nonmetricity, the torsion and the curvature:
\begin{eqnarray}
V &=& {\frac {1}{2\kappa c}}\Big\{a_0\eta_{\alpha\beta}\wedge R^{\alpha\beta}
-\,T^\alpha\wedge\sum_{I=1}^3 a_I\,{}^*({}^{(I)}T_\alpha)\nonumber\\
&& \qquad -\,Q_{\alpha\beta}\wedge \sum_{I=1}^{4}b_{I}\,^*({}^{(I)}Q^{\alpha\beta}) \nonumber\\
&& \qquad -\,2b_5({}^{(3)}Q_{\alpha\gamma}\wedge\vartheta^{\alpha})\wedge
{}^*({}^{(4)}Q^{\beta\gamma}\wedge\vartheta_{\beta})\nonumber\\
&& \qquad -\,2\vartheta^\alpha\wedge ^*\!T^\beta\wedge \sum_{I=1}^{3}c_{I}
\,{}^{(I+1)}Q_{\alpha\beta}\,\Bigr\}\nonumber\\
&& - \,{\frac 1{2\rho}}R^{\alpha\beta}\wedge {}^*\Bigl[\sum_{I=1}^6 w_I\,{}^{(I)}\!W_{\alpha\beta} 
 + \sum_{I=1}^5 z_I\,{}^{(I)}\!Z_{\alpha\beta} \nonumber\\
&& \qquad +\,v_1\,\vartheta_\alpha\wedge (e_\gamma\rfloor{}^{(5)}\!W^\gamma{}_\beta)
+ v_2\,\vartheta_\gamma\wedge (e_\alpha\rfloor{}^{(2)}\!Z^\gamma{}_\beta)\nonumber\\
&& \qquad +\,\sum_{I=3}^5 v_I\,\vartheta_\alpha\wedge (e_\gamma\rfloor{}^{(I)}\!Z^\gamma{}_\beta)\Bigr].\label{LRT}
\end{eqnarray}
We do not use topological invariants to simplify the Lagrangian. In this relation it is worthwhile to notice that the metric-affine Gauss-Bonnet term is not a boundary term in the presence of nonmetricity \cite{JanssenJC:2019}, and hence it cannot be used to eliminate quadratic curvature terms. For completeness, we included in (\ref{LRT}) the dimensionless constant $a_0$. This allows for the special case $a_0 = 0$ of the purely quadratic model without the Hilbert-Einstein linear term in the Lagrangian. In the Einstein-Cartan model, one puts $a_0 = 1$. Here we assume that the cosmological constant is zero. The analysis of a possibly nonvanishing cosmological term would require a different (more general) wave ansatz. 

The structure of the quadratic part of the general Lagrangian (\ref{LRT}) is determined by 27 dimensionless coupling constants: $a_1, a_2, a_3$, $b_1, \dots, b_5$, $c_1, c_2, c_3$, $w_1, \dots, w_6$, $z_1, \dots, z_5$, $v_1, \dots, v_5$. The coupling constant $\rho$ has the dimension of an inverse action: $[\rho] = [{\frac 1\hbar}]$. It is worthwhile to notice that this Lagrangian contains only parity-even invariants. For the most general case (to be analysed elsewhere) one should also take into account the parity-odd sector. 

The contribution of the curvature square terms in the Lagrangian (\ref{LRT}) to the gravitational field dynamics in the field equations is characterized by the parameter
\begin{equation}
\ell_\rho^2 := {\frac {\kappa c}{\rho}}.\label{lr}
\end{equation}
Since $[{\frac 1\rho}] = [\hbar]$, this new coupling parameter has the dimension of the area, $[\ell_\rho^2] = [\ell^2]$.

\section{Gravitational waves in MAG}\label{GW}

Let us now describe the plane wave ansatz in metric-affine gravity for the gravitational field potentials $(g_{\alpha\beta}, \vartheta^\alpha, \Gamma_\alpha{}^\beta)$. We will do it by extending the approach \cite{ppmag,yno:2017,BCO} in which the gravitational waves are patterned by the electromagnetic waves on a curved spacetime. 

As a first step, we divide the local coordinates into two groups: $x^i= (x^a, x^A)$, where $x^a = (x^0 = \sigma, x^1 = \rho)$ and $x^A = (x^2,x^3)$. Hereafter the indices from the beginning of the Latin alphabet $a,b,c... = 0,1$, whereas the capital Latin indices run $A,B,C... = 2,3$. 

To begin with, we fix the metric as the Minkowski tensor
\begin{equation}\label{gab}
g_{\alpha\beta} = \left(\begin{array}{crrr} 1 & 0 & 0 & 0\\ 0 & -1 & 0 & 0\\
0 & 0 & -1 & 0\\ 0 & 0 & 0 & -1\end{array}\right). 
\end{equation}
This can always be done by making use of the local general linear transformations of the coframe. Although it is common to use the so-called null (or semi-null) Minkowski metric for the discussion of the gravitational waves, however, throughout this paper we make use of the standard diagonal metric (\ref{gab}).

As the next step, we have to specify the ansatz for the coframe and the linear connection $(\vartheta^\alpha, \Gamma_\alpha{}^\beta)$. The coframe 1-form is chosen as
\begin{eqnarray}
\vartheta^{\widehat 0} &=& {\frac 12}(U + 1)d\sigma + {\frac 12}\,d\rho,\label{cof0}\\ 
\vartheta^{\widehat 1} &=& {\frac 12}(U - 1)d\sigma + {\frac 12}\,d\rho,\label{cof1}\\
\vartheta^{\widehat A} &=& dx^A,\qquad A = 2,3.\label{cof23}
\end{eqnarray}
Here $U = U(\sigma, x^A)$. As a result, the line element reads
\begin{equation}\label{ds_2}
ds^2 = g_{\alpha\beta}\vartheta^\alpha\vartheta^\beta = d\sigma d\rho + Ud\sigma^2 - \delta_{AB}dx^Adx^B.
\end{equation}

Following the analogy with the electromagnetism, we now introduce a crucial object: the wave 1-form $k$. We define the latter as 
\begin{equation}
k := d\sigma = \vartheta^{\widehat 0} - \vartheta^{\widehat 1}.\label{kdef}
\end{equation}
By construction, we have $k\wedge{}^\ast\!k = 0$. The wave covector $k_\alpha = e_\alpha\rfloor k$ then has (anholonomic) components $k_\alpha = (1, -1, 0, 0)$ and $k^\alpha = (1, 1, 0, 0)$. Hence, this is a null vector field, $k_\alpha k^\alpha = 0$. 

For the local Lorentz connection 1-form, we assume
\begin{equation}\label{conW}
\Gamma_\alpha{}^\beta = -\,k\left(k_\alpha V^\beta + k^\beta W_\alpha\right) + k_\alpha k^\beta\,u,
\end{equation}
where the two new vector variables are introduced: $W_\alpha = W_\alpha(\sigma, x^A)$, and $V^\alpha = V^\alpha(\sigma, x^A)$. The 1-form $u = u_\alpha(\sigma, x^A)\,\vartheta^\alpha$ is assumed to be orthogonal to the wave covector,
\begin{equation}
k\wedge{}^*\!u = 0,\qquad k^\alpha u_\alpha = 0.\label{ku}
\end{equation}
In addition, we assume the orthogonality 
\begin{equation}
k_\alpha W^\alpha = 0,\qquad k_\alpha V^\alpha = 0.\label{kW0}
\end{equation}
This is guaranteed if we choose 
\begin{equation}\label{Wa0}
W^\alpha = \begin{cases}W^a = 0,\qquad\qquad\qquad a = 0,1, \\
W^A = W^A(\sigma, x^B),\qquad A = 2,3.\end{cases}
\end{equation}
Here $W^2(\sigma, x^B)$ and $W^3(\sigma, x^B)$ are the two unknown functions. The same applies to $V^\alpha$:
\begin{equation}\label{Va0}
V^\alpha = \begin{cases}V^a = 0,\qquad\qquad\qquad a = 0,1, \\
V^A = V^A(\sigma, x^B),\qquad A = 2,3,\end{cases}
\end{equation}
and to the components of $u_\alpha$:
\begin{equation}\label{ua0}
u_\alpha = \begin{cases}u_a = 0,\qquad\qquad\qquad a = 0,1, \\
u_A = u_A(\sigma, x^B),\qquad A = 2,3,\end{cases}
\end{equation}

In other words, the ansatz for the MAG gauge potentials -- coframe (\ref{cof0})-(\ref{cof23}) and the linear connection (\ref{conW}) -- is described by 7 variables: $U = U(\sigma, x^B)$, $W^A = W^A(\sigma, x^B)$, $V^A = V^A(\sigma, x^B)$, and $u_A = u_A(\sigma, x^B)$. These functions determine wave's profile and their explicit form should be found from the gravitational field equations.

One immediately verifies that the wave 1-form is closed, and the wave covector is constant:
\begin{equation}
dk = 0,\qquad dk_\alpha = 0,\qquad Dk_\alpha = 0.\label{dk0}
\end{equation}
Taking this into account, we straightforwardly compute nonmetricity 1-form and the torsion and the curvature 2-forms:
\begin{eqnarray}
Q_{\alpha\beta} &=& -\,k\,(k_\alpha\!\pW_\beta + k_\beta\!\pW_\alpha) + 2k_\alpha k_\beta\,u,\label{nonW}\\
T^\alpha &=& -\,k\wedge k^\alpha\,\Theta,\label{torW}\\
R_\alpha{}^\beta &=& k\wedge\left(k_\alpha \underline{d}\,V^\beta
+ k^\beta \underline{d}\,W_\alpha\right) + k_\alpha k^\beta du,\label{curW}
\end{eqnarray}
where we introduced 
\begin{eqnarray}
\Theta &:=& {\frac 12}\,\underline{d}\,U + W_\alpha\vartheta^\alpha + u,\label{THW}\\
\pmW{\!}_\alpha &:=& W_\alpha \pm V_\alpha.\label{pmW}
\end{eqnarray}
The differential $\underline{d}$ acts in the transversal 2-space spanned by $x^A = (x^2, x^3)$:
\begin{equation}
 \underline{d} := \vartheta^Ae_A\rfloor d = dx^A\partial_A,\qquad A = 2,3.\label{ud}
\end{equation}
Although the geometry of the transversal 2-space spanned by $x^A = (x^2, x^3)$ is fairly simple, it is convenient to describe it explicitly. It is a flat Euclidean space with the volume 2-form $\underline{\eta} = {\frac 12}\eta_{AB}\vartheta^A\wedge\vartheta^B = dx^2\wedge dx^3$, where $\eta_{AB} = -\,\eta_{BA}$ is the 2-dimensional Levi-Civita tensor (with $\eta_{23} = 1$). The volume 4-form of the spacetime manifold then reads $\eta = \vartheta^{\widehat 0}\wedge \vartheta^{\widehat 1}\wedge\vartheta^{\widehat 2}\wedge\vartheta^{\widehat 3} = {\frac 12}k\wedge d\rho\wedge\underline{\eta}$. For the wave 1-form we find the remarkable relation 
\begin{equation}
{}^*k = -\,k\wedge\underline{\eta}.\label{dualk}
\end{equation}
We will denote the geometrical objects on the transversal 2-space by underlining them; for example, a 1-form $\underline{\phi} = \phi_A\vartheta^A$. The Hodge duality on this space is defined as usual via ${}^{\underline{*}}\vartheta_A = \underline{\eta}_A = e_A\rfloor \underline{\eta} = \eta_{AB}\vartheta^B$. With the help of (\ref{dualk}), we can verify
\begin{equation}
{}^*(k\wedge\underline{\phi}) = k\wedge{}^{\underline{*}}\underline{\phi}.\label{kphi}
\end{equation}

From (\ref{curW}) we immediately find
\begin{eqnarray}
W^{\alpha\beta} &=& R^{[\alpha\beta]} = -\,k\wedge k^{[\alpha}\mOm{\!}^{\beta]},\label{WW}\\
Z^{\alpha\beta} &=& R^{(\alpha\beta)} = k\wedge k^{(\alpha}\pOm{\!}^{\beta)} + k^\alpha k^\beta du,\label{ZW}
\end{eqnarray}
where we denoted
\begin{eqnarray}
\pmOm{\!}^\alpha := \underline{d}\pmW{\!}^\alpha. \label{OMW}
\end{eqnarray}
The new objects (\ref{THW}) and (\ref{OMW}) have the obvious properties:
\begin{equation}
k\wedge{}^*\Theta = 0,\quad k\wedge{}^*\pmOm{\!}^\alpha = 0,\quad 
k_\alpha\pmOm{\!}^\alpha = 0.\label{kTOM}
\end{equation}
In accordance with (\ref{Wa0})-(\ref{ua0}), we have explicitly: $\pmOm{\!}^a = 0$ ($a = 0,1$) and 
\begin{equation}
\Theta = \vartheta^A\!\left({\frac 12}\partial_AU - \delta_{AB}W^B + u_A\!\right),\
\!\pmOm{\!}^A = \vartheta^B\partial_B\!\pmW{\!}^A.\label{THOMa}
\end{equation}
Applying the transversal differential to (\ref{THW}), and making use of (\ref{OMW}), we find
\begin{equation}
\underline{d}\Theta = {\frac 12}(\pOm{\!}_\alpha + \mOm{\!}_\alpha)\wedge\vartheta^\alpha
+ \underline{d}u.\label{dTHW}
\end{equation}
In essence, this is equivalent to the Bianchi identity $DT^\alpha = R_\beta{}^\alpha\wedge\vartheta^\beta$ which is immediately checked by applying the covariant differential $D$ to (\ref{torW}) and using (\ref{curW}). It is worthwhile to notice that
\begin{equation}\label{du}
du = k\wedge\dot{u} + \underline{d}u,\qquad \dot{u} = (\partial_\sigma u_\alpha)\,\vartheta^\alpha.
\end{equation}

Let us discuss the properties of the torsion and the curvature for the wave ansatz (\ref{cof0})-(\ref{cof23}) and (\ref{conW}). To begin with, it is worthwhile to notice that the 2-forms of the gravitational gauge field strengths (\ref{torW}) and (\ref{curW}) have the same structure as the electromagnetic field strength of a plane wave, when $u = 0$. Indeed, we then have
\begin{equation}
Q_{\alpha\beta} = k\,q_{\alpha\beta},\quad T^\alpha = k\wedge a^\alpha,\quad R_\alpha{}^\beta = k\wedge a_\alpha{}^\beta,\label{poinW}
\end{equation}
where $q_{\alpha\beta} = -(k_\alpha\!\!\pW_\beta + k_\beta\!\!\pW_\alpha)$, $a^\alpha = -\, k^\alpha\Theta$ and $a_\alpha{}^\beta = k_\alpha \underline{d}\,V^\beta + k^\beta \underline{d}\,W_\alpha$ play the role of the gravitational ``polarization'' 1-forms, in complete analogy to the polarization 1-form $a$ in the electromagnetic plane wave field $F = k\wedge a$, which is orthogonal to the wave covector $k\wedge {}^*a = 0$. Similarly, the polarization 1-forms satisfy the orthogonality relations
\begin{eqnarray}
k\wedge{}^\ast a^\alpha = 0,\qquad k\wedge{}^\ast a_\alpha{}^\beta = 0,\label{aka1}\\
k^\alpha q_{\alpha\beta} = 0,\qquad k_\alpha a^\alpha = 0,\label{aka0}\\
k_\beta a_\alpha{}^\beta = 0,\qquad k^\alpha a_\alpha{}^\beta = 0.\label{aka2}
\end{eqnarray}
Clearly, the gravitational field strengths of a wave have the properties
\begin{eqnarray}
k\wedge{}^\ast\!Q_{\alpha\beta} = 0,\ k\wedge{}^\ast\!T^\alpha = 0,\ k\wedge{}^\ast\!R_\alpha{}^\beta = 0,\label{kTW}\\
k\wedge Q_{\alpha\beta} = 0,\ k\wedge T^\alpha = 0,\ k\wedge R_\alpha{}^\beta = 0,\label{kRW}\\
Q_{\alpha\beta}\wedge{}^\ast\!Q_{\rho\sigma} = 0,\ T^\alpha\wedge{}^\ast\!T^\beta = 0,\ R_\alpha{}^\beta\wedge{}^\ast\!R_\rho{}^\sigma = 0,\label{kTRW}
\end{eqnarray}
in complete analogy to the electromagnetic plane wave, $k\wedge F = 0$, $F\wedge {}^*F = 0$.

In addition, however, the gravitational field strengths satisfy 
\begin{equation}
k^\alpha Q_{\alpha\beta} = 0,\quad k_\alpha T^\alpha = 0,\quad k_\beta R_\alpha{}^\beta = 0,\quad k^\alpha R_\alpha{}^\beta = 0,\label{kTRW2}
\end{equation}
in view of (\ref{kW0}) and (\ref{kTOM}).

\subsection{Irreducible decomposition of gravitational field strengths}\label{irreps}

It is straightforward to find the irreducible parts of the nonmetricity, the torsion and the curvature.

Directly from (\ref{torW}), with an account of (\ref{kW0}), we find $e_\alpha\rfloor T^\alpha = 0$ and $\vartheta_\alpha\wedge T^\alpha = 0$. Hence the second (trace) and third (axial trace) irreducible parts of the torsion are trivial, ${}^{(2)}\!T^\alpha = 0$ and ${}^{(3)}\!T^\alpha = 0$, and 
\begin{equation}
{}^{(1)}\!T^\alpha = T^\alpha = -\,k\wedge k^\alpha\,\Theta.\label{torW1}
\end{equation}

In a similar way, we derive $e^\alpha\rfloor\!\aQ_{\alpha\beta} = 0$ and $Q_\alpha{}^\alpha = 0$ from (\ref{nonW}) and (\ref{kW0}). Hence the third and the fourth irreducible parts of the nonmetricity are trivial, ${}^{(3)}\!Q_{\alpha\beta} = 0$ and ${}^{(4)}\!Q_{\alpha\beta} = 0$, and we are left with
\begin{eqnarray}
{}^{(1)}\!Q_{\alpha\beta} &=& -\,{\frac 43}kk_{(\alpha}\!\pW_{\beta)}
- {\frac 23}k_\alpha k_\beta\!\pW_\gamma\!\vartheta^\gamma\nonumber\\
&& +\,{\frac 43}kk_{(\alpha}u_{\beta)} + {\frac 23}k_\alpha k_\beta\,u,\label{nonW1}\\
{}^{(2)}\!Q_{\alpha\beta} &=& -\,{\frac 23}kk_{(\alpha}\!\pW_{\beta)} 
+ {\frac 23}k_\alpha k_\beta\!\pW_\gamma\!\vartheta^\gamma\nonumber\\
&& -\,{\frac 43}kk_{(\alpha}u_{\beta)} + {\frac 43}k_\alpha k_\beta\,u.\label{nonW2}
\end{eqnarray}

Finally, the structure of the curvature $R^{\alpha\beta} = W^{\alpha\beta} + Z^{\alpha\beta}$ is as follows. Five irreducible pieces are trivial, ${}^{(3)}\!W^{\alpha\beta} = {}^{(5)}\!W^{\alpha\beta} = {}^{(6)}\!W^{\alpha\beta} = 0$, and ${}^{(3)}\!Z^{\alpha\beta} = {}^{(5)}\!Z^{\alpha\beta} = 0$, whereas for $I = 1,2,4$ we derive
\begin{eqnarray}
{}^{(I)}\!W^{\alpha\beta} &=& k\wedge {}^{(I)}\!\!\mOm{\!}^{[\alpha}k^{\beta]},\label{curW124}\\
{}^{(1)}\!Z^{\alpha\beta} &=& {\frac 12}k\wedge {}^{(1)}\!\!\pOm{\!}^{(\alpha}k^{\beta)}
+ {\frac 14}k^\alpha k^\beta\, \vartheta_\gamma\wedge\pOm{\!}^\gamma \nonumber\\
&& +\,{\frac 12}k\,\wedge\pOm{\!}^{(\alpha}k^{\beta)}\nonumber\\
&& +\,{\frac 12}k\wedge k^{(\alpha}e^{\beta)}\rfloor du + {\frac 12}k^\alpha k^\beta du,\label{ZW1}\\
{}^{(2)}\!Z^{\alpha\beta} &=& {\frac 12}k\wedge {}^{(2)}\!\!\pOm{\!}^{(\alpha}k^{\beta)}
-  {\frac 14}k^\alpha k^\beta\, \vartheta_\gamma\wedge\pOm{\!}^\gamma\nonumber\\
&& -\,{\frac 12}k\wedge k^{(\alpha}e^{\beta)}\rfloor du + {\frac 12}k^\alpha k^\beta du,\label{ZW2}\\
{}^{(4)}\!Z^{\alpha\beta} &=& {\frac 12}k\wedge {}^{(4)}\!\!\pOm{\!}^{(\alpha}k^{\beta)}.\label{ZW4}
\end{eqnarray}
These are constructed in terms of the irreducible parts
\begin{equation}\label{Omm}
\pmOm{\!}^\alpha = {}^{(1)}\!\!\pmOm{\!}^\alpha + {}^{(2)}\!\!\pmOm{\!}^\alpha + {}^{(4)}\!\!\pmOm{\!}^\alpha,
\end{equation}
which read explicitly: 
\begin{eqnarray}
{}^{(1)}\!\!\pmOm{\!}^\alpha &:=& {\frac 12}\left(\pmOm{\!}^\alpha - \vartheta^\alpha e_\beta\rfloor
\pmOm{\!}^\beta + \vartheta^\beta e^\alpha\rfloor\pmOm{\!}_\beta\right),\label{OM1}\\
{}^{(2)}\!\!\pmOm{\!}^\alpha &:=& {\frac 12}\left(\pmOm{\!}^\alpha - \vartheta^\beta e^\alpha\rfloor
\pmOm{\!}_\beta\right),\label{OM2}\\
{}^{(4)}\!\!\pmOm{\!}^\alpha &:=& {\frac 12}\,\vartheta^\alpha e_\beta\rfloor\pmOm{\!}^\beta.\label{OM4}
\end{eqnarray}
The transversal components of these objects are symmetric traceless part, skew-symmetric part and the trace of the $2\times 2$ matrix $\partial_B\!\pmW{\!}^A$, respectively. Using (\ref{THOMa}), we derive ${}^{(I)}\!\!\pmOm{\!}^A = {}^{(I)}\!\!\pmOm{\!}^A{}_B\,\vartheta^B$, with
\begin{eqnarray}
{}^{(1)}\!\!\pmOm{\!}^A{}_B &=& {\frac 12}\!\left(\!\partial_B\!\pmW{\!}^A + \partial^A\!\pmW{\!}_B
- \delta^A_B\,\partial_C\!\pmW{\!}^C\!\right)\!,\label{OMAB1}\\
{}^{(2)}\!\!\pmOm{\!}^A{}_B &=& {\frac 12}\!\left(\!\partial_B\!\pmW{\!}^A - \partial^A\!\pmW{\!}_B\!\right)\!,\label{OMAB2}\\
{}^{(4)}\!\!\pmOm{\!}^A{}_B &=& {\frac 12}\,\delta^A_B\,\partial_C\!\pmW{\!}^C.\label{OMAB4}
\end{eqnarray}

One can demonstrate the following properties of these 1-forms:
\begin{eqnarray}\label{vOM}
\vartheta_\alpha\wedge{}^{(1)}\!\Omega^\alpha = 0,\quad \vartheta_\alpha\wedge{}^{(2)}\!\Omega^\alpha 
= \vartheta_\alpha\wedge\Omega^\alpha,\\
\vartheta_\alpha\wedge{}^{(4)}\!\Omega^\alpha = 0,\quad
e_\alpha\rfloor{}^{(1)}\!\Omega^\alpha = -\,e_\alpha\rfloor\Omega^\alpha,\\ 
e_\alpha\rfloor{}^{(2)}\!\Omega^\alpha = 0,\quad e_\alpha\rfloor{}^{(4)}\!\Omega^\alpha = 
2e_\alpha\rfloor\Omega^\alpha,\label{eOM}\\
k_\alpha{}^{(1)}\!\Omega^\alpha = -\,{\frac 12}\,k\,e_\alpha\rfloor\Omega^\alpha,\quad
k_\alpha{}^{(2)}\!\Omega^\alpha = 0,\label{kOM1}\\
k_\alpha{}^{(4)}\!\Omega^\alpha = {\frac 12}\,k\,e_\alpha\rfloor\Omega^\alpha,
\quad k\wedge{}^*{}^{(2)}\!\Omega^\alpha = 0,\label{kOM4}\\
k\wedge{}^*{}^{(1)}\!\Omega^\alpha = -\, k\wedge{}^*{}^{(4)}\!\Omega^\alpha = 
-\,{\frac 12}\,k^\alpha\,\vartheta_\beta\wedge{}^*\Omega^\beta.\label{kdOM}
\end{eqnarray}
To make formulas more compact, we omit the $^{(\pm)}$ labels because the above properties hold for both types of $\Omega$'s.

It is worthwhile to notice that although some of the irreducible parts (\ref{OM1})-(\ref{OM4}) are not orthogonal to the wave covector, in view of (\ref{kOM1}) and (\ref{kOM4}), all irreducible parts of the curvature and nonmetricity satisfy
\begin{equation}
k_\alpha {}^{(I)}\!W^{\alpha\beta} = 0,\quad k_\alpha {}^{(I)}\!Z^{\alpha\beta} = 0,\quad
k_\alpha {}^{(I)}\!Q^{\alpha\beta} = 0,\label{kWZQ}
\end{equation}
in full agreement with (\ref{kTRW2}).

\section{Explicit field equations}\label{MAGE}

For the Lagrangian (\ref{LRT}) from the definitions (\ref{Mab}) and (\ref{Ha}), we find explicitly the gravitational field momenta
\begin{eqnarray}
M^{\alpha\beta} &=& {\frac {2}{\kappa c}}\,m^{\alpha\beta},\qquad
H_\alpha = {\frac {1}{\kappa c}}\,h_\alpha,\label{MH}\\
m^{\alpha\beta} &=& \sum_{I=1}^4 b_I\,{}^*({}^{(I)}Q^{\alpha\beta})\nonumber\\
&& +\,b_5\Bigl[ \vartheta^{(\alpha}\wedge e^{\beta)}\rfloor {}^*Q - {\frac 14}g^{\alpha\beta}\,{}^*(\Lambda + 3Q)\Bigr]\nonumber\\
&& -\,c_1\vartheta^{(\alpha}\wedge {}^*T^{\beta)} 
+ {\frac {c_1 + c_2}{3}}\vartheta^{(\alpha}\wedge e^{\beta)}\rfloor {}^*T\nonumber\\
&& +\, {\frac {2c_1 - c_2 - c_3}{4}}g^{\alpha\beta}\,{}^*T,\label{LMab}\\
h_{\alpha} &=& \sum_{I=1}^3 a_I\,{}^*({}^{(I)}T_\alpha)\nonumber\\
&& +\,\sum_{I=1}^3 c_I\,{}^*(\vartheta^\beta\wedge {}^{(I+1)}Q_{\alpha\beta}).\label{LHa}
\end{eqnarray}
For the MAG wave ansatz, these are reduced to 
\begin{eqnarray}
m^{\alpha\beta} &=& {}^*\!\Bigl[b_1{}^{(1)}Q^{\alpha\beta} 
+ b_2{}^{(2)}Q^{\alpha\beta} + c_1e^{(\alpha}\rfloor T^{\beta)}\Bigr],\label{GWMab}\\
h_{\alpha} &=& {}^*\!\Bigl[a_1\,T_\alpha
- c_1\,{}^{(2)}Q_{\alpha\beta}\wedge\vartheta^\beta\Bigr].\label{GWHa}
\end{eqnarray}
A direct computation from the definition (\ref{Hab}) for the Lagrangian (\ref{LRT}) yields, by inserting the MAG wave ansatz:
\begin{equation}
H^\alpha{}_\beta = -\,{\frac {a_0}{2\kappa c}}\,\eta^\alpha{}_\beta
+ {\frac 1\rho}\,h^\alpha{}_\beta,\label{HH}
\end{equation}
where $h_{\alpha\beta} = h_{[\alpha\beta]} + h_{(\alpha\beta)}$ explicitly read
\begin{widetext}
\begin{eqnarray}
h_{[\alpha\beta]} &=& {}^*\Bigl[w_1{}^{(1)}\!W_{\alpha\beta} + w_2{}^{(2)}\!W_{\alpha\beta}
  + w_4{}^{(4)}\!W_{\alpha\beta}
  + {\frac {v_2}{2}}\,\vartheta_\gamma\wedge e_{[\alpha}\rfloor {}^{(2)}\!Z^\gamma{}_{\beta]} + 
{\frac {v_4}{2}}\,\vartheta_{[\alpha}\wedge e_{|\gamma|}\rfloor {}^{(4)}\!Z^\gamma{}_{\beta]}\Bigr],\label{HAGW}\\
h_{(\alpha\beta)} &=& {}^*\Bigl[z_1{}^{(1)}\!Z_{\alpha\beta}
+ (z_2 - v_2){}^{(2)}\!Z_{\alpha\beta} + (z_4 + 2v_4){}^{(4)}\!Z_{\alpha\beta} 
+ {\frac {v_2}{2}}\,\vartheta_\gamma\wedge e_{(\alpha}\rfloor {}^{(2)}\!W^\gamma{}_{\beta)} + 
{\frac {v_4}{2}}\,\vartheta_{(\alpha}\wedge e_{|\gamma|}\rfloor {}^{(4)}\!W^\gamma{}_{\beta)}\Bigr].\label{HSGW}
\end{eqnarray}
Splitting the Lagrangian (\ref{LRT}) into the linear Hilbert-Einstein term and the purely quadratic part, $V = {\frac {a_0}{2\kappa c}}\,\eta_{\alpha\beta}\wedge R^{\alpha\beta} + {\frac {1}{\kappa c}}\,V^{(q)}$, we derive from (\ref{Ea})
\begin{equation}\label{Ea1}
E_\alpha = {\frac {a_0}{2\kappa c}}\,\eta_{\alpha\beta\gamma}\wedge R^{\beta\gamma}
+ {\frac {1}{\kappa c}}\,q_\alpha,
\end{equation}
where we introduced
\begin{eqnarray}
q_{\alpha} := e_{\alpha}\rfloor V^{(q)} + (e_{\alpha}\rfloor T^{\beta})\wedge h_{\beta}
+ \ell_\rho^2\,(e_{\alpha}\rfloor R_{\beta}{}^{\gamma})\wedge h^{\beta}{}_{\gamma}
 + (e_{\alpha}\rfloor Q_{\beta\gamma})\,m^{\beta\gamma}.\label{qa}
\end{eqnarray}
\end{widetext}
With an account of (\ref{MH}), (\ref{HH}), and (\ref{Ea1}), the MAG field equations (\ref{dVt}) and (\ref{dVG}) in vacuum (assuming the vanishing matter sources $\Sigma_\alpha = 0$ and $\Delta^\alpha{}_\beta = 0$) are recast into the following form:
\begin{eqnarray}
{\frac {a_0}{2}}\,\eta_{\alpha\beta\gamma}\wedge R^{\beta\gamma} + q_\alpha - Dh_\alpha = 0,\label{1st}\\
{\frac {a_0}{2}}\,D\eta^\alpha{}_\beta - \vartheta^\alpha\wedge h_\beta
- 2m^\alpha{}_\beta - \ell_\rho^2\,Dh^\alpha{}_\beta = 0.\label{2nd}
\end{eqnarray}
The first term in (\ref{2nd}) is straightforwardly evaluated
\begin{eqnarray}
D\eta^\alpha{}_\beta = D(g^{\alpha\gamma}\eta_{\gamma\beta}) = \eta^\alpha{}_{\beta\gamma}\wedge T^\gamma\nonumber\\
+ \,Q^{\alpha\gamma}\wedge\eta_{\gamma\beta} - 2Q\wedge\eta^\alpha{}_\beta.\label{deta}
\end{eqnarray}

In view of (\ref{kWZQ}), we verify the orthogonality properties for (\ref{MH}), (\ref{HAGW}) and (\ref{HSGW}):
\begin{equation}
k_\alpha m^{\alpha\beta} = 0,\quad k^\alpha h_\alpha = 0,\quad k^\alpha h_{[\alpha\beta]} = 0,
\quad k^\alpha h_{(\alpha\beta)} = 0.\label{kmhh}
\end{equation}
As a result, for (\ref{qa}) we derive
\begin{equation}
q_\alpha = 0\label{qaW}
\end{equation}
for the MAG wave field strengths (\ref{nonW})-(\ref{curW}). Moreover, substituting (\ref{curW124})-(\ref{ZW4}) into (\ref{HAGW}) and (\ref{HSGW}), we recast the latter into
\begin{eqnarray}
h_{[\alpha\beta]} &=& {}^*(k\wedge{\stackrel {(w)}\Sigma}{}_{[\alpha})\,k_{\beta]},\label{HAGW1}\\
h_{(\alpha\beta)} &=& {}^*(k\wedge{\stackrel {(z)}\Sigma}{}_{(\alpha})\,k_{\beta)}
+ {\frac 14}\,k_\alpha k_\beta\,{}^*\!\Sigma,\label{HSGW1}
\end{eqnarray}
where
\begin{eqnarray}
{\stackrel {(w)}\Sigma}{}_\alpha &:=& w_1{}^{(1)}\!\!\mOm{\!}_\alpha + w_2{}^{(2)}\!\!\mOm{\!}_\alpha
+ w_4{}^{(4)}\!\!\mOm{\!}_\alpha \nonumber\\
&& - \,{\frac {v_2}2}{}^{(2)}\!\!\pOm{\!}_\alpha + {\frac {v_4}2}{}^{(4)}\!\!\pOm{\!}_\alpha
+ {\frac {v_2}2}e_\alpha\rfloor\underline{d}u,\label{Lw}\\
{\stackrel {(z)}\Sigma}{}_\alpha &:=& z_1{}^{(1)}\!\!\pOm{\!}_\alpha
+ \,{\frac {z_1 + z_2 - v_2}2}{}^{(2)}\!\!\pOm{\!}_\alpha\nonumber\\
&& + {\frac {z_1 + z_4 + 2v_4}2}{}^{(4)}\!\!\pOm{\!}_\alpha \nonumber\\
&& -\,{\frac {v_2}2}{}^{(2)}\!\!\mOm{\!}_\alpha + {\frac {v_4}2}{}^{(4)}\!\!\mOm{\!}_\alpha\nonumber\\
&& +\,{\frac {z_1 - z_2 + v_2}2}\,e_\alpha\rfloor\underline{d}u,\label{Lz}\\
\Sigma &:=& \vartheta^\gamma\wedge [(z_1 - z_2 + v_2)\pOm{\!}_\gamma + v_2\mOm{\!}_\gamma]\nonumber\\
&& + \,4z_1k\wedge\dot{u} + 2(z_1 + z_2 - v_2)\underline{d}u.\label{L}
\end{eqnarray}
In a similar way, substituting (\ref{torW}) and (\ref{nonW1})-(\ref{nonW2}) into (\ref{GWMab}) and (\ref{GWHa}), we recast the latter into
\begin{eqnarray}
m^{\alpha\beta} &=& {}^*(k\wedge\mu^{(\alpha})\,k^{\beta)} - k^\alpha k^\beta\,{}^*\!\mu,\label{GWMab1}\\
h_{\alpha} &=& -\,k_\alpha\,{}^*\!\Bigl[k\wedge(a_1\,\Theta - c_1\,{\stackrel {(+)}W}{}_\beta
\vartheta^\beta - 2c_1\,u)\Bigr],\label{GWHa1}
\end{eqnarray}
where
\begin{eqnarray}
\mu_\alpha &:=& c_1\,e_\alpha\rfloor\Theta - {\frac {4b_1 + 2b_2}{3}}{\stackrel {(+)}W}{}_\alpha
+ {\frac {4(b_1 - b_2)}{3}}u_\alpha,\label{mua}\\
\mu &:=& c_1\,\Theta + {\frac {2b_1 - 2b_2}{3}}{\stackrel {(+)}W}{}_\beta\vartheta^\beta
- {\frac {2b_1 + 4b_2}{3}}\,u.\label{mu}
\end{eqnarray}
In view of (\ref{kphi}), the structure of the 2-forms (\ref{HAGW1}), (\ref{HSGW1}) and (\ref{GWHa1}) guarantees that
\begin{equation}
\Gamma_\alpha{}^\beta\wedge h_\beta = 0,\quad \Gamma_\alpha{}^\beta\wedge h_{[\beta\gamma]} = 0,\quad
\Gamma_\alpha{}^\beta\wedge h_{(\beta\gamma)} = 0,\label{Ghh}
\end{equation}
for the connection (\ref{conW}), and therefore the covariant derivatives are reduced to the ordinary ones:
\begin{equation}
Dh_\alpha = dh_\alpha,\qquad Dh^\alpha{}_\beta = dh^\alpha{}_\beta.\label{Dhdh}
\end{equation}
With an account of (\ref{qaW}), (\ref{deta}) and (\ref{Dhdh}), we can recast the MAG field equations (\ref{1st}) and (\ref{2nd}) into the final form
\begin{eqnarray}
{\frac {a_0}{2}}\,\eta_{\alpha\beta\gamma}\wedge R^{\beta\gamma} - dh_\alpha &=& 0,\label{1stF}\\
{\frac {a_0}{2}}\,(\eta_{\alpha\beta\gamma}\wedge T^\gamma + Q_{[\alpha}{}^\gamma\wedge\eta_{|\gamma|\beta]})&&\nonumber\\
- \,\vartheta_{[\alpha}\wedge h_{\beta]} - \ell_\rho^2\,dh_{[\alpha\beta]} &=& 0,\label{2nd1}\\
{\frac {a_0}{2}}\,Q_{(\alpha}{}^\gamma\wedge\eta_{|\gamma|\beta)} - \vartheta_{(\alpha}\wedge h_{\beta)}&&\nonumber\\
- \,2m_{\alpha\beta} - \ell_\rho^2\,dh_{(\alpha\beta)} &=& 0.\label{2nd2}
\end{eqnarray}
After making use of (\ref{Dhdh}), we have lowered the upper index and split the resulting equation into the symmetric and skew-symmetric equations to derive (\ref{2nd1}) and (\ref{2nd2}).

It thus remains to plug (\ref{HAGW1})-(\ref{mu}) into the field equations (\ref{1stF})-(\ref{2nd2}) and solve the system of coupled equations. It is worthwhile to notice the remarkable fact that the final system is a set of {\it linear} differential equations for the unknown functions $U = U(\sigma, x^B)$, $W^A = W^A(\sigma, x^B)$, $V^A = V^A(\sigma, x^B)$, and $u_A = u_A(\sigma, x^B)$.

\subsection{First field equation}\label{mag1}

Making use of (\ref{WW}) we find
\begin{eqnarray}
\eta_{\alpha\beta\gamma}\wedge R^{\beta\gamma} &=& \eta_{\alpha\beta\gamma}\wedge W^{\beta\gamma} = 
\eta_{\alpha\beta\gamma}\wedge k\,\wedge \mOm{\!}^\beta k^\gamma \nonumber\\
&=& k_\alpha\,{}^*k\,e_\beta\rfloor \mOm{\!}^\beta.\label{HE}
\end{eqnarray}
On the other hand, we have, see (\ref{GWHa1}), $h_{\alpha} = -\,k_\alpha\,{}^*(k\wedge\Xi)$ with
the {\it transversal} 1-form
\begin{equation}
\Xi := a_1\,\Theta - c_1\,{\stackrel {(+)}W}{}_\beta \vartheta^\beta - 2c_1\,u,\label{XI}
\end{equation}
and in view of (\ref{kphi}), we evaluate the exterior differential 
\begin{equation}
dh_\alpha = k_\alpha k\wedge \underline{d}\,{}^{\underline{*}}\Xi.\label{dh}
\end{equation}
It is worthwhile to note that although $\Xi$ depends on $\sigma$, we have for the total differential
\begin{equation}
d\,{}^{\underline{*}}\Xi = d\sigma\wedge\partial_\sigma({}^{\underline{*}}\Xi)
+ \underline{d}\,{}^{\underline{*}}\Xi,\label{dXi}
\end{equation}
and thus the first term drops out from (\ref{dh}) since $d\sigma = k$, cf. (\ref{kdef}).

Finally, by using the dual (\ref{dualk}), we recast the first MAG field equation (\ref{1stF}) into
\begin{equation}
-\,k_\alpha k\wedge \Bigl\{{\frac {a_0}{2}}\,\underline{\eta}\,e_\beta\rfloor\!\mOm{\!}^\beta
+ \underline{d}\,{}^{\underline{*}}\Xi\Bigr\} = 0.\label{1stF1}
\end{equation}

\subsection{Second field equation}\label{mag2}

Since $\vartheta_\alpha\wedge h_\beta = {}^*(e_\alpha\rfloor{}^*h_\beta)$, a direct computation yields,
with the help of (\ref{nonW}) and (\ref{torW}):
\begin{eqnarray}
{\frac {a_0}{2}}\,\eta_{\alpha\beta\gamma}\wedge T^\gamma &=& a_0\,{}^*k\,e_{[\alpha}\rfloor\Theta\,k_{\beta]},\label{etaT}\\
{\frac {a_0}{2}}Q_\alpha{}^\gamma\wedge\eta_{\gamma\beta} - \vartheta_\alpha\wedge h_\beta
&=& {}^*(k\nu_\alpha) k_\beta - k_\alpha k_\beta {}^*\nu\nonumber\\
&& +\,a_0\,{}^*k\,k_\alpha u_\beta,\label{Qeta}
\end{eqnarray}
where
\begin{eqnarray}
\nu_\alpha &:=& a_1\,e_\alpha\rfloor\Theta - ({\frac {a_0}{2}} + c_1)\,{\stackrel {(+)}W}{}_\alpha
- 2c_1 u_\alpha,\label{nua}\\
\nu &:=& a_1\,\Theta + ({\frac {a_0}{2}} - c_1)\,{\stackrel {(+)}W}{}_\beta\vartheta^\beta
- 2c_1\,u.\label{nu}
\end{eqnarray}
As a result, for the second MAG field equations (\ref{2nd1}) and (\ref{2nd2}) we find 
\begin{eqnarray}
{\frac {a_0}{2}}\,(\eta_{\alpha\beta\gamma}\wedge T^\gamma + Q_{[\alpha}{}^\gamma\wedge\eta_{|\gamma|\beta]})
- \,\vartheta_{[\alpha}\wedge h_{\beta]} \nonumber\\
= {}^*k\left(a_0\,e_{[\alpha}\rfloor\Theta + \nu_{[\alpha} - a_0u_{[\alpha}\right)k_{\beta]},\label{etaT1}\\
{\frac {a_0}{2}}\,Q_{(\alpha}{}^\gamma\wedge\eta_{|\gamma|\beta)} - \vartheta_{(\alpha}\wedge h_{\beta)}\nonumber\\
= {}^*k\,(\nu_{(\alpha} + a_0u_{(\alpha})\, k_{\beta)} - k_\alpha k_\beta {}^*\nu.\label{Qeta1}
\end{eqnarray}

Finally, the differentials of (\ref{HAGW1}) and (\ref{HSGW1}) read
\begin{eqnarray}
dh_{[\alpha\beta]} &=& -\,k\wedge \underline{d}\,{}^{\underline{*}}
{\stackrel {(w)}\Sigma}{}_{[\alpha}\,k_{\beta]},\label{dha}\\
dh_{(\alpha\beta)} &=& d\,{}^*(k\wedge{\stackrel {(z)}\Sigma}{}_{(\alpha})\,k_{\beta)}
+ {\frac 14}\,k_\alpha k_\beta\,d\,{}^*\!\Sigma,\label{dhS}
\end{eqnarray}
where we used (\ref{kphi}) once again. Note that in (\ref{dhS}) we have the full differential, unlike the transversal one in (\ref{dha}), and evaluation of the latter is somewhat nontrivial. 

When writing down the field equations, we have to specialize to the subsets of the indices: $\alpha = (a, A)$, with lower case Latin indices $a = 0,1$ and the upper case indices $A = 2,3$.

As a preliminary step, we recall that
\begin{equation}
k_A = 0,\label{kA0}
\end{equation}
and notice that
\begin{eqnarray}
{}^{(1)}\!\pmOm{\!}^a &=& -\,{\frac 12}\,\vartheta^a\,e_\beta\rfloor\!\pmOm{\!}^\beta,\label{pmOm1}\\
{}^{(2)}\!\pmOm{\!}^a &=& 0,\label{pmOm2}\\
{}^{(4)}\!\pmOm{\!}^a &=& {\frac 12}\,\vartheta^a\,e_\beta\rfloor\!\pmOm{\!}^\beta,\label{pmOm4}
\end{eqnarray}
whereas ${}^{(I)}\!\!\pmOm{\!}^A$ are purely transversal for all $I = 1,2,4$.

As a result, from (\ref{Lw}) and (\ref{Lz}) we conclude that ${\stackrel {(w)}\Sigma}{}_A$ and ${\stackrel {(z)}\Sigma}{}_A$ are purely transversal, while
\begin{eqnarray}
{\stackrel {(w)}\Sigma}{}_a = {\frac 14}\,\vartheta_a{\stackrel {(w)}\varphi},\qquad
{\stackrel {(z)}\Sigma}{}_a = {\frac 14}\,\vartheta_a{\stackrel {(z)}\varphi},\label{Lwza}
\end{eqnarray}
where
\begin{eqnarray}
{\stackrel {(w)}\varphi} &:=& 2(w_4 - w_1)\,e_\beta\rfloor\!\mOm{\!}^\beta
+ v_4\,e_\beta\rfloor\!\pOm{\!}^\beta,\label{phiw}\\
{\stackrel {(z)}\varphi} &:=& (z_4 - z_1 + 2v_4)\,e_\beta\rfloor\!\pOm{\!}^\beta
+ v_4\,e_\beta\rfloor\!\mOm{\!}^\beta.\label{phiz}
\end{eqnarray}

Then, for (\ref{HAGW1}) we find
\begin{eqnarray}
h_{[ab]} &=& 0,\qquad h_{[AB]} = 0,\label{habA1}\\
h_{[Ab]} &=& {\frac 12}\,{}^*(k\wedge{\stackrel {(w)}\Sigma}{}_A)k_b = 
{\frac 12}\,k\wedge{}^{\underline{*}}{\stackrel {(w)}\Sigma}{}_A\,k_b.\label{habA2}
\end{eqnarray}
The structure of (\ref{HSGW1}) is more nontrivial:
\begin{eqnarray}
h_{(ab)} &=& {}^*(k\wedge{\stackrel {(z)}\Sigma}{}_{(a})\,k_{b)} + {\frac 14}\,k_a k_b\,{}^*\Sigma\nonumber\\
&=& {\frac 14}\,k_a k_b\Bigl\{{}^*\Sigma - {\stackrel {(z)}\varphi}\,\underline{\eta}\Bigr\},\label{habS1}\\
h_{(Ab)} &=& {\frac 12}\,{}^*(k\wedge{\stackrel {(z)}\Sigma}{}_A)\,k_b = 
{\frac 12}\,k\wedge{}^{\underline{*}}{\stackrel {(z)}\Sigma}{}_A\,k_b.\label{habS2}\\
h_{(AB)} &=& 0.\label{habS3}
\end{eqnarray}
Here we used (\ref{Lwza}) and  
\begin{equation}
{}^*(k\wedge\vartheta_a) = e_a\rfloor{}^*k = -\,k_a\,\underline{\eta}.\label{kaeta}
\end{equation}

After these preparations, we are in a position to write down the second MAG field equations. The skew-symmetric part (\ref{2nd1}) has only only ``$[Ab]$'' nontrivial component which reads explicitly
\begin{equation}
k\wedge\Bigl\{\underline{\eta}\left(a_0\,e_A\rfloor\Theta + \nu_A - a_0u_A\right) - \ell_\rho^2
\,\underline{d}\,{}^{\underline{*}}{\stackrel {(w)}\Sigma}{}_A\Bigr\} k_b = 0.\label{2nd10}
\end{equation}
However, the symmetric part (\ref{2nd2}) encompasses two nontrivial components, ``$(ab)$'' and
``$(Ab)$'', respectively:
\begin{eqnarray}
k_ak_b\Bigl\{ {}^*(2\mu - \nu) - {\frac {\ell_\rho^2}4}\,d\,({}^*\Sigma - {\stackrel {(z)}\varphi}
\,\underline{\eta})\Bigr\} &=& 0,\label{2nd21}\\
k\wedge\Bigl\{\underline{\eta}\left(\nu_A - 2\mu_A + a_0\,u_A\right) - \ell_\rho^2
\,\underline{d}\,{}^{\underline{*}}{\stackrel {(z)}\Sigma}{}_A\Bigr\} k_b &=& 0.\label{2nd22}
\end{eqnarray}

\subsection{Explicit MAG field equations in components}\label{magc}

To begin with, let us fix the notation. Namely, we choose the {\it original} position of indices for the vector objects as upper position for $W^\alpha$ and $V^\alpha$, and lower position for $u_\alpha$. In other words, the basic variables will be chosen as
\begin{equation}
W^A,\qquad V^A,\qquad u_A.\label{WVu0}
\end{equation}
Then we denote these objects with the indices moved to a different position as
\begin{equation}
\underline{W}_A := \delta_{AB}W^B,\quad \underline{V}_A := \delta_{AB}V^B,
\quad \underline{u}^A := \delta^{AB}u_B.\label{WVu1}
\end{equation}
In addition, we denote the differential operator
\begin{equation}
\underline{\partial}^A := \delta^{AB}\partial_B.\label{dA}
\end{equation}
This convention is extremely important when we recast the 4-dimensional expressions in the formulas of the sections above into the 2-dimensional transversal ones. In particular, one should be always careful with $\partial^\alpha$, $W_\alpha$, $V_\alpha$, and $u^\alpha$, when we specialize to $\alpha = A$, since then
\begin{equation}
\partial^A = -\,\underline{\partial}^A,\quad W_A = -\,\underline{W}_A,\quad 
V_A = -\,\underline{V}_A,\quad u^A = -\,\underline{u}^A.\label{dWVu}
\end{equation}

As a preliminary step, we observe the Hodge duals
\begin{equation}\label{hodge}
{}^*\underline{\eta} = \vartheta^{\hat{0}}\wedge\vartheta^{\hat{1}},\qquad {}^*\underline{\phi}
= \vartheta^{\hat{0}}\wedge\vartheta^{\hat{1}}\wedge{}^{\underline{*}}\underline{\phi},
\end{equation}
where the latter is true for any transversal 1-form $\underline{\phi}$. It is also worthwhile to notice that $\vartheta^{\hat{0}}\wedge\vartheta^{\hat{1}} = {\frac 12}\,d\sigma\wedge d\rho$ and thus the exterior differential vanishes: $d\left(\vartheta^{\hat{0}}\wedge\vartheta^{\hat{1}}\right) = 0$.

Let us analyse (\ref{2nd21}). Since from (\ref{L}) we have
\begin{equation}\label{Sigma}
\Sigma = \chi\,\underline{\eta} + k\wedge\xi,
\end{equation}
then by making use of (\ref{hodge}) and (\ref{kphi}) we obtain for the Hodge dual ${}^*\Sigma = \chi\,\vartheta^{\hat{0}}\wedge\vartheta^{\hat{1}} + k\wedge {}^{\underline{*}}\xi$, and thus
\begin{equation}\label{dSigma}
d\,{}^*\Sigma = \vartheta^{\hat{0}}\wedge\vartheta^{\hat{1}}\wedge\underline{d}\chi
- k\wedge \underline{d}\,{}^{\underline{*}}\xi.
\end{equation}
Comparing (\ref{Sigma}) with (\ref{L}), we explicitly have for the transversal 1-form $\xi = 4z_1\dot{u}$, and 
\begin{widetext}
\begin{equation}
\chi = (z_1 - z_2 + 2v_2)\,\eta^{AB}\partial_A\underline{W}_B + 
(z_1 - z_2)\,\eta^{AB}\partial_A\underline{V}_B
+ 2(z_1 + z_2 - v_2)\,\eta^{AB}\partial_Au_B.\label{chi}
\end{equation}
As a result, with the help of (\ref{hodge}) and (\ref{dSigma}) we recast (\ref{2nd21}) into
\begin{equation}
\vartheta^{\hat{0}}\wedge\vartheta^{\hat{1}}\wedge\Bigl\{ 2{}^{\underline{*}}\mu - {}^{\underline{*}}\nu
- {\frac {\ell_\rho^2}4}\,\underline{d}\,\chi\Bigr\} + {\frac {\ell_\rho^2}4}\,k\wedge\Bigl\{
\underline{\eta}\,\partial_\sigma{\stackrel {(z)}\varphi} + \underline{d}\,{}^{\underline{*}}\xi
\Bigr\} = 0.\label{2nd21A}
\end{equation}
Therefore, in components this yields two equations:
\begin{eqnarray}
{\frac {2c_1 - a_1}{2}}\partial_AU + \Bigl[ {\frac {a_0 + 2a_1}{2}} - 3c_1 - {\frac {4b_1 - 4b_2}{3}}
\Bigr]\underline{W}_A + \Bigl[ {\frac {a_0}{2}} - c_1 - {\frac {4b_1 - 4b_2}{3}} \Bigr]
\underline{V}_A  + \Bigl[4c_1 - a_1 - {\frac {4b_1 + 8b_2}{3}}\Bigr] u_A\nonumber\\
+ \,{\frac {\ell_\rho^2}{4}}\,\eta_{AB}\,\underline{\partial}^B\Bigl[(z_1 - z_2 + 2v_2)\,
\eta^{CD}\partial_C\underline{W}_D + (z_1 - z_2)\,\eta^{CD}\partial_C\underline{V}_D
+ 2(z_1 + z_2 - v_2)\,\eta^{CD}\partial_Cu_D \Bigr] = 0,\label{2nd211}\\ 
\partial_\sigma\Bigl[(z_4 - z_1 + 3v_4)\partial_AW^A + (z_4 - z_1 + v_4)\partial_AV^A 
- 4z_1\partial_A\underline{u}^A\Bigr] = 0.\label{2nd212}
\end{eqnarray}
In addition, the two more field equations (\ref{2nd10}) and (\ref{2nd22}) read in components:
\begin{eqnarray}
{\frac {a_0 + a_1}{2}}\partial_AU + \Bigl[ c_1 - {\frac {a_0 + 2a_1}{2}}\Bigr]\underline{W}_A
+ \Bigl[ {\frac {a_0}{2}} + c_1\Bigr]\underline{V}_A  + \Bigl[a_1 - 2c_1\Bigr] u_A\nonumber\\
- \,{\frac {\ell_\rho^2}{4}}\Bigl[2w_1\Delta\underline{W}_A - 2w_1\Delta\underline{V}_A 
+ (2w_4 + v_4)\partial_A\partial_BW^B + (-2w_4 + v_4)\partial_A\partial_BV^B \Bigr]\nonumber\\
- \,{\frac {\ell_\rho^2}{4}}\,\eta_{AB}\,\underline{\partial}^B\Bigl[(- 2w_2 + v_2)\,
\eta^{CD}\partial_C\underline{W}_D + (2w_2 + v_2)\,\eta^{CD}\partial_C\underline{V}_D
- 2v_2\,\eta^{CD}\partial_Cu_D \Bigr] & = 0,\label{2nd11}
\end{eqnarray}
\begin{eqnarray}
{\frac {a_1 - 2c_1}{2}}\partial_AU + \Bigl[ {\frac {a_0}{2}} - a_1 + 3c_1 - {\frac {8b_1 + 4b_2}{3}}
\Bigr]\underline{W}_A + \Bigl[ {\frac {a_0}{2}} + c_1 - {\frac {8b_1 + 4b_2}{3}}
\Bigr]\underline{V}_A  + \Bigl[a_0 + a_1 - 4c_1 - {\frac {8b_1 - 8b_2}{3}}\Bigr] u_A\nonumber\\
- \,{\frac {\ell_\rho^2}{4}}\Bigl[2z_1\Delta\underline{W}_A + 2z_1\Delta\underline{V}_A 
+ (z_1 + z_4 + 3v_4)\partial_A\partial_BW^B + (z_1 + z_4 + v_4)\partial_A\partial_BV^B \Bigr]\qquad\nonumber\\
- \,{\frac {\ell_\rho^2}{4}}\,\eta_{AB}\,\underline{\partial}^B\Bigl[(- z_1 - z_2 + 2v_2)\,
\eta^{CD}\partial_C\underline{W}_D - (z_1 + z_2)\,\eta^{CD}\partial_C\underline{V}_D
- 2(z_1 - z_2 + v_2)\,\eta^{CD}\partial_Cu_D \Bigr] = 0.\nonumber\\ \label{2nd12}
\end{eqnarray}
Finally, to make the system complete, we write explicitly the first MAG field equation (\ref{1stF1}):
\begin{equation}
-\,{\frac {a_1}{2}}\,\Delta U + \Bigl[{\frac {a_0}{2}} - c_1 + a_1\Bigr]\partial_AW^A
- \Bigl[{\frac {a_0}{2}} + c_1\Bigr]\partial_AV^A - \Bigl[a_1 - 2c_1\Bigr]\partial_A\underline{u}^A
 = 0.\label{1stF2}
\end{equation}
The total number of equations (\ref{2nd211}), (\ref{2nd11})-(\ref{1stF2}) is 7 which is equal to the number unknown variables $U, W^A, V^A, u_A$, so we expect that one can find the latter as functions of transversal coordinates $x^A$. An additional equation (\ref{2nd212}) does not make the system over-determined, since it merely fixes the dependence on $\sigma$. 
\end{widetext}

\section{Solving field equations}\label{FE}

We are now in a position to solve the field equations. The system (\ref{2nd211})-(\ref{1stF2}) always admits a nontrivial solution for the arbitrary quadratic MAG model with any choice of coupling constants. There are some interesting special cases.

\subsection{Riemannian gravitational waves}

The nonmetricity (\ref{nonW}) and the torsion (\ref{torW}) vanish when $u = 0$, $\pW{\!}^\alpha = 0$, and $\Theta = 0$ which is realized for 
\begin{equation}
W^A = -\,V^A = {\frac 12}\delta^{AB}\partial_BU.\label{notorW}
\end{equation}
Substituting this into (\ref{2nd211})-(\ref{1stF2}), we find that (\ref{2nd211}) 
is identically satisfied, the first MAG equation (\ref{1stF2}) reduces to
\begin{equation}
{\frac {a_0}{2}}\,\Delta\,U = 0,\label{notW1}
\end{equation}
whereas (\ref{2nd212})-(\ref{2nd12}) reduce to 
\begin{eqnarray}
v_4\,\partial_\sigma\,\Delta\,U &=& 0,\label{notW0}\\
-\,2(w_1 + w_4)\,{\frac {\ell_\rho^2}{4}}\,\partial_A\,\Delta\,U &=& 0,\label{notW2}\\
-\,v_4\,{\frac {\ell_\rho^2}{4}}\,\partial_A\,\Delta\,U &=& 0.\label{notW3}
\end{eqnarray}
Accordingly, we conclude that the well-known plane wave solution of GR with the function $U$ satisfying the Laplace equation is an exact solution of the generic quadratic MAG model. This is consistent with our earlier results on the torsion-free solutions in Poincar\'e gauge theory \cite{selected,Obukhov:1989,yno:2019}.

Moreover, the Riemannian wave (\ref{notorW})-(\ref{notW1}) represents a general solution for the purely torsion + nonmetricity quadratic class of MAG models, since this is the only configuration admitted by the system (\ref{2nd211})-(\ref{1stF2}) for $w_I = 0$, $z_I = 0$, $v_I = 0$. This is true generically when the curvature square terms are absent, with an exception of a special choice of the coupling constants \cite{obukhov:1997}:
\begin{eqnarray}
-\,a_1 = {\frac {a_2}{2}} = 2a_3 = 2c_1 = - c_2 = - c_3 = a_0,\label{acex}\\
4b_1 = 2b_2 = - 8b_3 = {\frac {8b_4}{3}} = 2b_5  = a_0.\label{bex}
\end{eqnarray}

\subsection{Teleparallel gravitational waves}

Quite generally, the space of distant parallelism (or the teleparallel space) is defined by the condition of zero curvature, $R_\alpha{}^\beta = 0$. In the framework of MAG, the two other gravitational field strengths, torsion and nonmetricity, are nontrivial, and therefore the general teleparallel gravitational waves are characterized by the superposition of propagating torsion and nonmetricity waves, provided the curvature is trivial.

The curvature (\ref{curW}) vanishes when $\underline{d}W^\alpha = 0$ and $\underline{d}V^\alpha = 0$, i.e., both $W^\alpha = W^\alpha(\sigma)$ and $V^\alpha = V^\alpha(\sigma)$ are independent of the transversal coordinates, and in addition $du = 0$. The latter means that the components of $u$ do not depend on $\sigma$, and
\begin{equation}\label{udU}
u_A = {\frac 12}\,\partial_A\,{\mathcal U}
\end{equation}
is a gradient of a potential ${\mathcal U} = {\mathcal U}(x^A)$.

Then (\ref{2nd212}) is identically fulfilled, and the three equations (\ref{2nd211}), (\ref{2nd11}), and (\ref{2nd12}) after a long but straightforward derivation are recast into an equivalent algebraic system
\begin{eqnarray}
(a_0 - 4b_1)\,\Phi_A = 0.\label{e1}\\
3(a_0 + 2c_1)\Theta_A + 2(a_0 + 2b_2)\,\Psi_A = 0,\label{e2}\\
2(a_0 + a_1)\,\Theta_A + (a_0 + 2c_1)\,\Psi_A = 0,\label{e3}
\end{eqnarray}
where we denoted 
\begin{eqnarray}
\Theta_A=e_A\rfloor\Theta &=& {\frac 12}\partial_AU - \underline{W}_A + u_A,\label{thA}\\
\Phi_A &:=& \underline{W}_A + \underline{V}_A + u_A,\label{phA}\\
\Psi_A &:=& \underline{W}_A + \underline{V}_A - 2u_A.\label{psA}
\end{eqnarray}

In addition, the equation (\ref{1stF2}) is reduced to
\begin{equation}
a_1\,\Delta\,U + (a_1 - 2c_1)\,\Delta\,{\mathcal U} = 0.\label{teleW0}
\end{equation}

Writing the system (\ref{e1})-(\ref{e3}) in the matrix form,
\begin{equation}
\left(\begin{array}{ccc} 0 & (a_0 - 4b_1) & 0 \\
2(a_0 + a_1) & 0 & (a_0 + 2c_1) \\
3(a_0 + 2c_1) & 0 & 2(a_0 + 2b_2) \end{array}\right)
\left(\begin{array}{c}\Theta_A \\ \Phi_A \\ \Psi_A\end{array}\right) =  0,\label{matrix}
\end{equation}
we conclude that a nontrivial solution exists when the determinant is zero:
\begin{equation}\label{det}
(a_0 - 4b_1)\left[3(a_0 + 2c_1)^2 - 4(a_0 + a_1)(a_0 + 2b_2)\right] = 0.
\end{equation}
This imposes the restriction on the coupling constants of the general Lagrangian (\ref{LRT}) and determines the class of MAG models which admit the teleparallel gravitational waves. 

The explanation of the condition (\ref{det}) is as follows. If the determinant is not zero, then the system (\ref{e1})-(\ref{e3}) yields a trivial solution $\Theta_A = \Phi_A = \Psi_A = 0$ which means that all the gravitational field strengths are zero: the curvature $R_\alpha{}^\beta = 0$, the nonmetricity $Q_{\alpha\beta} = 0$, and the torsion $T^\alpha =  0$. Therefore such a solution describes the flat Minkowski spacetime.

\subsection{Standard teleparallel waves: no nonmetricity}

The nonmetricity (\ref{nonW}) vanishes when $u = 0$ and $\pW{\!}^\alpha = 0$, i.e. $W^\alpha = - V^\alpha$. This means that $\Phi_A = \Psi_A = 0$, and the system (\ref{e1})-(\ref{e3}) is greatly simplified. As a result, such a solution only exists in a class of quadratic models restricted by the conditions on the coupling constants
\begin{equation}
a_0 + a_1 = 0,\qquad c_1 + {\frac {a_0}{2}} = 0.\label{teleW}
\end{equation}
The system (\ref{2nd211})-(\ref{1stF2}) then reduces to
\begin{equation}
{\frac {a_1}{2}}\,\Delta\,U = 0,\label{teleW1}
\end{equation}
Accordingly, the metric structure turns out to be the same for the Riemannian (no torsion and nonmetricity) and for the teleparallel gravitational wave solutions.
\bigskip

\subsection{Symmetric teleparallel waves: no torsion}

Symmetric teleparallel geometry is characterized by the vanishing curvature and torsion, along with a nontrivial nonmetricity \cite{Nester:1999,Adak:2006a,Adak:2006b,Jimenez:2018,Conroy:2018,Hohmann:2018,Hohmann:2019}. This case arises when $du = 0$ and both $W^\alpha = W^\alpha(\sigma)$ and $V^\alpha = V^\alpha(\sigma)$ are independent of the transversal coordinates, whereas $\Theta = 0$. The latter means that, by making use of (\ref{thA}) and (\ref{udU}), we have
\begin{equation}
\partial_A(U + {\mathcal U}) - 2\underline{W}_A = 0.\label{udUW}
\end{equation}
As a result, the algebraic system (\ref{e1})-(\ref{e3}) is simplified to 
\begin{equation}\label{Xu1}
(a_0 - 4b_1)\,\Phi_A = 0,\ (a_0 + 2b_2)\,\Psi_A = 0,\ (a_0 + 2c_1)\,\Psi_A = 0,
\end{equation}
whereas (\ref{1stF2}) reduces to
\begin{equation}
-\,c_1\,\Delta\,U = 0.\label{teleS1}
\end{equation}

Consequently, nontrivial symmetric teleparallel wave solutions exist for the class of MAG models restricted by the conditions on the coupling constants
\begin{equation}
a_0 - 4b_1 = 0,\quad a_0 + 2b_2 = 0,\quad c_1 + {\frac {a_0}{2}} = 0.\label{teleS}
\end{equation} 
Otherwise, solutions reduce to the flat Minkowski spacetime. The conditions (\ref{teleS}) allow for the general solution with both $\Psi_A$ and $\Phi_A$ nonvanishing. Special symmetric teleparallel wave solutions with $\Phi_A = 0$ or $\Psi_A = 0$ exist under the milder conditions when one drops one of the restriction in (\ref{teleS}).

\subsection{General MAG gravitational waves}

The torsion-free ansatz (\ref{notorW}) can be generalized to 
\begin{eqnarray}
W^A &=& {\frac 12}\left(\delta^{AB}\partial_B{\mathcal W} + \eta^{AB}\partial_B\overline{\mathcal W}\right),\label{pW}\\
V^A &=& {\frac 12}\left(\delta^{AB}\partial_B{\mathcal V} + \eta^{AB}\partial_B\overline{\mathcal V}\right),\label{pV}\\
u_A &=& {\frac 12}\left(\partial_A{\mathcal U} + \eta_{AB}\,\underline{\partial}^B\,\overline{\mathcal U}\right),\label{pu}
\end{eqnarray}
Physically, six new variables ${\mathcal W}, {\mathcal V}, {\mathcal U}$ and $\overline{\mathcal W}, \overline{\mathcal V}, \overline{\mathcal U}$ are analogs of the well-known Hertz potentials in classical electrodynamics. The overline denotes the three {\it parity-odd} variables $\overline{\mathcal W}, \overline{\mathcal V}, \overline{\mathcal U}$ to distinguish them from the {\it parity-even} variables $U, {\mathcal W}, {\mathcal V}, {\mathcal U}$.

\begin{widetext}
Substituting this into (\ref{2nd211})-(\ref{1stF2}), we derive
\begin{equation}
(2c_1 - a_1)\,U + \Bigl[ {\frac {a_0 + 2a_1}{2}} - 3c_1 - {\frac {4b_1 - 4b_2}{3}}\Bigr]\,{\mathcal W}
+ \Bigl[ {\frac {a_0}{2}} - c_1 - {\frac {4b_1 - 4b_2}{3}} \Bigr]\,{\mathcal V}
+ \Bigl[4c_1 - a_1 - {\frac {4b_1 + 8b_2}{3}}\Bigr]\,{\mathcal U} = 0,\label{eq1}
\end{equation}
\begin{eqnarray}
\Bigl[ {\frac {a_0 + 2a_1}{2}} - 3c_1 - {\frac {4b_1 - 4b_2}{3}}\Bigr]\,\overline{\mathcal W}
+ \Bigl[ {\frac {a_0}{2}} - c_1 - {\frac {4b_1 - 4b_2}{3}} \Bigr]\,\overline{\mathcal V}
+ \Bigl[4c_1 - a_1 - {\frac {4b_1 + 8b_2}{3}}\Bigr]\,\overline{\mathcal U}&& \nonumber\\
-\,{\frac {\ell_\rho^2}{4}}\,\Bigl[(z_1 - z_2 + 2v_2)\,\Delta\,\overline{\mathcal W}
+ (z_1 - z_2)\,\Delta\,\overline{\mathcal V}
+ 2(z_1 + z_2 - v_2)\,\Delta\,\overline{\mathcal U} \Bigr] &=& 0,\label{eq2}\\ 
\partial_\sigma\,\Bigl[(z_4 - z_1 + 3v_4)\,\Delta\,{\mathcal W} + (z_4 - z_1 + v_4)\,\Delta\,{\mathcal V}
- 4z_1\,\Delta\,{\mathcal U}\Bigr] &=& 0,\label{eq3}
\end{eqnarray}
\begin{eqnarray}
(a_0 + a_1)\,U + \Bigl[ c_1 - {\frac {a_0 + 2a_1}{2}}\Bigr]\,{\mathcal W}
+ \Bigl[ {\frac {a_0}{2}} + c_1\Bigr]\,{\mathcal V}  + \Bigl[a_1 - 2c_1\Bigr]\,{\mathcal U}&&\nonumber\\ 
- \,{\frac {\ell_\rho^2}{4}}\,\Bigl[(2w_1 + 2w_4 + v_4)\,\Delta\,{\mathcal W}
- (2w_1 + 2w_4 - v_4)\,\Delta\,{\mathcal V}\Bigr] &=& 0,\label{eq4}\\
\Bigl[ c_1 - {\frac {a_0 + 2a_1}{2}}\Bigr]\,\overline{\mathcal W} + \Bigl[ {\frac {a_0}{2}} + c_1\Bigr]
\,\overline{\mathcal V} + \Bigl[a_1 - 2c_1\Bigr]\,\overline{\mathcal U}&&\nonumber\\ 
- \,{\frac {\ell_\rho^2}{4}}\,\Bigl[(2w_1 + 2w_2 - v_2)\,\Delta\,\overline{\mathcal W}
- (2w_1 + 2w_2 + v_2)\,\Delta\,\overline{\mathcal V} + 2v_2\,\Delta\,\overline{\mathcal U}\Bigr] &=& 0,\label{eq5}
\end{eqnarray}
\begin{eqnarray}
(a_1 - 2c_1)\,U + \Bigl[ {\frac {a_0}{2}} - a_1 + 3c_1 - {\frac {8b_1 + 4b_2}{3}}\Bigr]\,{\mathcal W}
+ \Bigl[ {\frac {a_0}{2}} + c_1 - {\frac {8b_1 + 4b_2}{3}}\Bigr]\,{\mathcal V}&&\nonumber\\
+ \Bigl[a_0 + a_1 - 4c_1 - {\frac {8b_1 - 8b_2}{3}}\Bigr]\,{\mathcal U}
- \,{\frac {\ell_\rho^2}{4}}\Bigl[(3z_1 + z_4 + 3v_4)\,\Delta\,{\mathcal W}  
+ (3z_1 + z_4 + v_4)\,\Delta\,{\mathcal V}\Bigr] &=& 0,\label{eq6}\\
\Bigl[ {\frac {a_0}{2}} - a_1 + 3c_1 - {\frac {8b_1 + 4b_2}{3}}\Bigr]\,\overline{\mathcal W}
+ \Bigl[ {\frac {a_0}{2}} + c_1 - {\frac {8b_1 + 4b_2}{3}}\Bigr]\,\overline{\mathcal V}
+ \Bigl[a_0 + a_1 - 4c_1 - {\frac {8b_1 - 8b_2}{3}}\Bigr]\,\overline{\mathcal U}&&\nonumber\\
- \,{\frac {\ell_\rho^2}{4}}\,\Bigl[(3z_1 + z_2 - 2v_2)\,\Delta\,\overline{\mathcal W} + (3z_1 + z_2)
\,\Delta\,\overline{\mathcal V} + 2(z_1 - z_2 + v_2)\,\Delta\,\overline{\mathcal U}\Bigr] &=& 0,\label{eq7}
\end{eqnarray}
\begin{equation}
-\,a_1\,\Delta U + \Bigl[{\frac {a_0}{2}} - c_1 + a_1\Bigr]\,\Delta\,{\mathcal W}
- \Bigl[{\frac {a_0}{2}} + c_1\Bigr]\,\Delta\,{\mathcal V}
- \Bigl[a_1 - 2c_1\Bigr]\,\Delta\,{\mathcal U}  = 0.\label{eq8}
\end{equation}
\end{widetext}

The analysis of this system is considerably simplified by a convenient choice of variables. The key to this is discovered when we substitute (\ref{pW})-(\ref{pu}) into (\ref{thA})-(\ref{psA}), which yields
\begin{eqnarray}
\Theta_A &=& {\frac 12}\left(\partial_A{\mathcal X}_1 + \eta_{AB}\,\underline{\partial}^B
\,\overline{\mathcal X}_1\right),\label{thA1}\\
\Phi_A &=& {\frac 12}\left(\partial_A{\mathcal X}_2 + \eta_{AB}\,\underline{\partial}^B
\,\overline{\mathcal X}_2\right),\label{psA1}\\
\Psi_A &=& {\frac 12}\left(\partial_A{\mathcal X}_3 + \eta_{AB}\,\underline{\partial}^B
\,\overline{\mathcal X}_3\right),\label{phiA1}
\end{eqnarray}
where
\begin{eqnarray}
{\mathcal X}_1 = U - {\mathcal W} + {\mathcal U},&\quad& \overline{\mathcal X}_1 =
- \overline{\mathcal W} + \overline{\mathcal U},\label{X1}\\
{\mathcal X}_2 = {\mathcal W} + {\mathcal V} + {\mathcal U},&\quad& \overline{\mathcal X}_2 =
\overline{\mathcal W} + \overline{\mathcal V} + \overline{\mathcal U},\label{X2}\\
{\mathcal X}_3 = {\mathcal W} + {\mathcal V} - 2{\mathcal U},&\quad& \overline{\mathcal X}_3 =
\overline{\mathcal W} + \overline{\mathcal V} - 2\overline{\mathcal U}.\label{X3}
\end{eqnarray}
We choose (\ref{X1})-(\ref{X3}) as the new set of variables, to which we add one more:
\begin{equation}
{\mathcal X}_0 = {\mathcal W} - {\mathcal V}.\label{X0}
\end{equation}
For the inverse relations we derive
\begin{eqnarray}
U &=& {\frac 12}{\mathcal X}_0 + {\mathcal X}_1 + {\frac 12}{\mathcal X}_3,\label{UX}\\
{\mathcal W} &=& {\frac 12}{\mathcal X}_0 + {\frac 13}{\mathcal X}_2 + {\frac 16}{\mathcal X}_3,\label{WX}\\
{\mathcal V} &=& - \,{\frac 12}{\mathcal X}_0 + {\frac 13}{\mathcal X}_2 + {\frac 16}{\mathcal X}_3,\label{VX}\\
{\mathcal U} &=& {\frac 13}{\mathcal X}_2 - {\frac 13}{\mathcal X}_3,\label{uX}
\end{eqnarray}
and similarly we find
\begin{eqnarray}
\overline{\mathcal W} &=& -\,\overline{\mathcal X}_1 + {\frac 13}\overline{\mathcal X}_2 - {\frac 13}\overline{\mathcal X}_3,\label{oWX}\\
\overline{\mathcal V} &=& \overline{\mathcal X}_1 + {\frac 13}\overline{\mathcal X}_2 + {\frac 23}\overline{\mathcal X}_3,\label{oVX}\\
\overline{\mathcal U} &=& {\frac 13}\overline{\mathcal X}_2 - {\frac 13}\overline{\mathcal X}_3.\label{ouX}
\end{eqnarray}

Substituting (\ref{UX})-(\ref{uX}) into the field equations, we recast (\ref{eq1}), (\ref{eq4}), (\ref{eq6}), (\ref{eq8}), (\ref{eq3}), respectively, into
\begin{widetext}
\begin{eqnarray}
(2c_1 - a_1)\,{\mathcal X}_1 + {\frac 13}(a_0 - 4b_1)\,{\mathcal X}_2 + 
\Bigl[ - \,{\frac {a_0}{2}} - c_1 + {\frac 23}(a_0 + 2b_2)\Bigr]\,{\mathcal X}_3 &=& 0,\label{Xeq1}\\
(a_0 + a_1)\,{\mathcal X}_1 + \Bigl({\frac {a_0}{2}} + c_1\Bigr)\,{\mathcal X}_3
- \,{\frac {\ell_\rho^2}{4}}\,\Bigl[2(w_1 + w_4)\,\Delta\,{\mathcal X}_0 + {\frac 23}\,v_4
\,\Delta\,{\mathcal X}_2 + {\frac 13}\,v_4\,\Delta\,{\mathcal X}_3 \Bigr] &=& 0,\label{Xeq4} \\
(a_1 - 2c_1)\,{\mathcal X}_1 + {\frac 23}(a_0 - 4b_1)\,{\mathcal X}_2 + 
\Bigl[ {\frac {a_0}{2}} + c_1 - {\frac 23}(a_0 + 2b_2)\Bigr]\,{\mathcal X}_3 &&\nonumber\\
- \,{\frac {\ell_\rho^2}{4}}\Bigl[ v_4\,\Delta\,{\mathcal X}_0 + {\frac 23}
\,(3z_1 + z_4 + 2v_4)\,\Delta\,{\mathcal X}_2 + {\frac 13}\,(3z_1 + z_4 + 2v_4)
\,\Delta\,{\mathcal X}_3\Bigr] &=& 0,\label{Xeq6}\\
{\frac {a_0}{2}}\,\Delta\,{\mathcal X}_0 - a_1\,\Delta\,{\mathcal X}_1 - c_1\,\Delta\,{\mathcal X}_3 
&=& 0,\label{Xeq8}\\
\partial_\sigma\Bigl\{v_4\,\Delta\,{\mathcal X}{}_0 + {\frac 23}(z_4 - 3z_1 + 2v_4)\,\Delta\,{\mathcal X}{}_2
+ {\frac 13}(z_4 + 3z_1 + 2v_4)\,\Delta\,{\mathcal X}{}_3\Bigr\} &=& 0.\label{Xeq3}
\end{eqnarray}
Similarly, substituting (\ref{oWX})-(\ref{ouX}) into the field equations, we recast (\ref{eq2}), (\ref{eq5}), and (\ref{eq7}), respectively, into
\begin{eqnarray}
(2c_1 - a_1)\,\overline{\mathcal X}_1 + {\frac 13}(a_0 - 4b_1)\,\overline{\mathcal X}_2 + 
\Bigl[ -\,{\frac {a_0}{2}} - c_1 + {\frac 23}(a_0 + 2b_2)\Bigr]\,\overline{\mathcal X}_3&& \nonumber\\
-\,{\frac {\ell_\rho^2}{4}}\,\Bigl[ -2v_2\,\Delta\,\overline{\mathcal X}_1 + {\frac 43}\,z_1
\,\Delta\,\overline{\mathcal X}_2 - {\frac 13}\,(z_1 + 3z_2)\,\Delta\,\overline{\mathcal X}_3
\Bigr] &=& 0,\label{Xeq2}\\ 
(a_0 + a_1)\,\overline{\mathcal X}_1 + \Bigl({\frac {a_0}{2}} + c_1\Bigr)\,\overline{\mathcal X}_3
- \,{\frac {\ell_\rho^2}{4}}\,\Bigl[-\,4(w_1 + w_2)\,\Delta\,\overline{\mathcal X}_1
- (2w_1 + 2w_2 + v_2)\,\Delta\,\overline{\mathcal X}_3 \Bigr] &=& 0,\label{Xeq5}\\
(a_1 - 2c_1)\,\overline{\mathcal X}_1 + {\frac 23}(a_0 - 4b_1)\,\overline{\mathcal X}_2 + 
\Bigl[ {\frac {a_0}{2}} + c_1 - {\frac 23}(a_0 + 2b_2)\Bigr]\,\overline{\mathcal X}_3 &&\nonumber\\
-\,{\frac {\ell_\rho^2}{4}}\,\Bigl[ 2v_2\,\Delta\,\overline{\mathcal X}_1 + {\frac 83}\,z_1
\,\Delta\,\overline{\mathcal X}_2 + {\frac 13}\,(z_1 + 3z_2)\,\Delta\,\overline{\mathcal X}_3
\Bigr] &=& 0.\label{Xeq7}
\end{eqnarray}

It is remarkable that the parity-even (\ref{Xeq1})-(\ref{Xeq3}) and the parity-odd (\ref{Xeq2})-(\ref{Xeq7}) sectors are completely decoupled.

Both are second order differential systems with constant coefficients, for which solutions are sought in the form
\begin{equation}\label{XX}
{\mathcal X}_I = {\mathcal X}_I^{(0)}(\sigma)\,e^{i\,q_A\,x^A},\quad I = 0,1,2,3,\qquad
\overline{\mathcal X}_J = \overline{\mathcal X}{}_J^{(0)}(\sigma)\,e^{i\,\overline{q}_A\,x^A},\quad J = 1,2,3,
\end{equation}
where  $q_A$ and  $\overline{q}_A$ are not necessarily equal.

{\it Parity-even sector}. Substituting the ansatz (\ref{XX}) into (\ref{Xeq1})-(\ref{Xeq8}) we obtain the algebraic system for the amplitudes ${\mathcal X}_I^{(0)}$. In matrix form, the latter reads as
\renewcommand\arraystretch{1.5}
\begin{equation}
\left(\begin{array}{cccc} -{\frac {a_0}{2}} & a_1 & 0 & c_1 \\
0 & 2c_1 - a_1 & {\frac 13}(a_0 - 4b_1) & - {\frac {a_0}{2}} - c_1 + {\frac 23}(a_0 + 2b_2) \\
2(w_1 + w_4){\mathcal Q}^2 & a_0 + a_1 & {\frac 23}v_4{\mathcal Q}^2 &
{\frac {a_0}{2}} + c_1 + {\frac 13}v_4{\mathcal Q}^2 \\
{\mathcal Q}^2v_4 & 0 & a_0 - 4b_1 + {\frac 23}{\mathcal Q}^2\Lambda_0 &
 {\frac 13}{\mathcal Q}^2\Lambda_0 \end{array}\right)
\left(\begin{array}{c}{\mathcal X}_0^{(0)} \\ {\mathcal X}_1^{(0)} \\
{\mathcal X}_2^{(0)} \\ {\mathcal X}_3^{(0)}\end{array}\right) =  0.\label{algX}
\end{equation}
\renewcommand\arraystretch{1}
Here we denoted ${\mathcal Q}^2  :=  {\frac {\ell^2}{4}q_Aq_B\delta^{AB}}$, and $\Lambda_0 := 3z_1 + z_4 + 2v_4$. Nontrivial solutions exist when the determinant of the $4\times 4$ matrix in (\ref{algX}) vanishes. Interestingly, despite being $4\times 4$, the matrix has a very special structure that yields just a quadratic equation for ${\mathcal Q}^2$. This equation determines the eigenvalues of the two propagating massive wave modes.

It is worthwhile to notice the existence of a special solution of (\ref{Xeq1})-(\ref{Xeq3}): ${\mathcal X}_2 = 0$ and ${\mathcal X}_3  = 0$. This means that ${\mathcal U} = 0$ and ${\mathcal W} = -\, {\mathcal V}$. The remaining system for ${\mathcal X}_0$ and ${\mathcal X}_1$ can be recast into
\begin{equation}
v_4\,\Delta {\mathcal X}_0 = 0,\qquad (2c_1 - a_1)\,{\mathcal X}_1 = 0, \qquad 
-\frac{a_0}{2}\Delta{\mathcal X}_0 +a_1\,\Delta{\mathcal X}_1 = 0,\qquad
(a_0+a_1)\,{\mathcal X}_1 - \frac{\ell_\rho^2}{2}(w_1+w_4)\Delta{\mathcal X}_0 = 0.\label{X2X3zero}
\end{equation}
Recalling the definitions (\ref{X1})-(\ref{X0}), and assuming $v_4\neq 0$ and $a_1\neq 2c_1$, we find $\Delta{\mathcal X}_0 = 0 ={\mathcal X}_1$ and the four equations (\ref{X2X3zero}) are automatically fulfilled. We thus conclude that this solution describes the massless graviton mode, for which the 4 potentials satisfy
\begin{equation}\label{mass0}
\Delta U = 0,\qquad {\mathcal W} = -\,{\mathcal V} = U,\qquad {\mathcal U} = 0.
\end{equation}

{\it Parity-odd sector}. A peculiar property of the system  (\ref{Xeq2})-(\ref{Xeq7}) is that the variable $\overline{\mathcal X}_2$ is decoupled from the pair $\overline{\mathcal X}_1, \overline{\mathcal X}_3$. Indeed, the sum of (\ref{Xeq2}) and (\ref{Xeq7}) yields an equation for $\overline{\mathcal X}_2$,
\begin{equation}\label{DX2}
(a_0 - 4b_1)\,\overline{\mathcal X}_2 - \ell_\rho^2\,z_1\,\Delta\,\overline{\mathcal X}_2 = 0,
\end{equation}
whereas by taking the sum of (\ref{Xeq2}) and (\ref{Xeq5}) we derive
\begin{equation}
(a_0 + 2c_1)\,\overline{\mathcal X}_1 + {\frac 23}(a_0 + 2b_2)\,\overline{\mathcal X}_3 
-\,{\frac {\ell_\rho^2}{4}}\,\Biggl\{ - 2\left[2w_1 + 2w_2 + v_2\right]\,\Delta
\,\overline{\mathcal X}_1 - \Bigl[2w_1 + 2w_2 + v_2 + {\frac 13}\,(z_1 + 3z_2)\Bigr]
\,\Delta\,\overline{\mathcal X}_3\Biggr\} = 0,\label{Xeq2a}
\end{equation}
which in combination with (\ref{Xeq5}) determines $\overline{\mathcal X}_1$ and $\overline{\mathcal X}_3$.

Substituting the ansatz (\ref{XX}) into (\ref{DX2}), (\ref{Xeq2a}), and (\ref{Xeq5}), we obtain the algebraic system for the amplitudes $\overline{\mathcal X}{}_I^{(0)}$. In matrix form, the latter reads as
\renewcommand\arraystretch{1.5}
\begin{equation}
\left(\begin{array}{ccc} 0 & a_0 - 4b_1 + 4z_1\overline{\mathcal Q}{}^2 & 0 \\
a_0 + 2c_1 - 2\overline{\mathcal Q}{}^2\Lambda_2 & 0 & {\frac 23}(a_0 + 2b_2)
- \overline{\mathcal Q}{}^2(\Lambda_2 + \Lambda_3)\\
a_0 + a_1 - \overline{\mathcal Q}{}^2\Lambda_1 & 0 & {\frac {a_0}{2}} + c_1
- \overline{\mathcal Q}{}^2\Lambda_2\end{array}\right)
\left(\begin{array}{c}\overline{\mathcal X}{}_1^{(0)} \\
\overline{\mathcal X}{}_2^{(0)} \\ \overline{\mathcal X}{}_3^{(0)}\end{array}\right) =  0.\label{algXo}
\end{equation}
\renewcommand\arraystretch{1}
Here we denoted $\overline{\mathcal Q}{}^2 := {\frac {\ell^2}{4}\overline{q}_A\overline{q}_B\delta^{AB}}$, and $\Lambda_1 := 4(w_1 + w_2)$, $\Lambda_2 := 2(w_1 + w_2) + v_2$, $\Lambda_3 := {\frac 13}(z_1 + 3z_2)$.

The system (\ref{algXo}) shows that there are three propagating parity-odd modes which are determined by
\begin{equation}
a_0 - 4b_1 + 4z_1\overline{\mathcal Q}{}^2 = 0,\qquad {\mathcal A}\overline{\mathcal Q}{}^4 +
 {\mathcal B}\overline{\mathcal Q}{}^2 +  {\mathcal C} = 0,\label{oddQ}
\end{equation}
where we denoted the combinations of the coupling constants
\begin{eqnarray}
{\mathcal A} &:=& 2\Lambda_2^2 + \Lambda_1(\Lambda_2 + \Lambda_3),\label{oddA}\\
{\mathcal B} &:=& -\,4\left({\frac {a_0}{2}} + c_1\right)\Lambda_2 + (a_0 + a_1)(\Lambda_2 + \Lambda_3) 
- {\frac 23}(a_0 + 2b_2)\Lambda_1,\label{oddB}\\
{\mathcal C} &:=& 2\left({\frac {a_0}{2}} + c_1\right)^2 - {\frac 23}(a_0 + 2b_2)(a_0 + a_1).\label{oddC}
\end{eqnarray}

The parity-odd amplitudes $\overline{\mathcal X}{}_I^{(0)} = \overline{\mathcal X}{}_I^{(0)}(\sigma)$,  $J = 1,2,3$, are arbitrary functions of $\sigma$. However, the field equation (\ref{Xeq3}) imposes a relation between the three parity-even amplitudes,
\begin{equation}\label{Xeq30}
v_4\,\partial_\sigma{\mathcal X}{}_0^{(0)} + {\frac 23}(z_4 - 3z_1 + 2v_4)\,\partial_\sigma
{\mathcal X}{}_2^{(0)} + {\frac 13}(z_4 + 3z_1 + 2v_4)\,\partial_\sigma{\mathcal X}{}_3^{(0)} = 0,
\end{equation}
whereas ${\mathcal X}_1^{(0)} = {\mathcal X}_1^{(0)}(\sigma)$ depends arbitrarily on $\sigma$.
\end{widetext}

\subsection{``Pseudo-instanton'' solutions}

Vassiliev \cite{vas1} considered the class of models in which the Lagrangian depends only on the curvature (with $a_1 = a_2 = a_3 = 0$, $b_1 = \dots = b_5 = 0$, $c_1 = c_2 = c_3 = 0$)\footnote{To be precise, Vassiliev's Lagrangian \cite{vas1} did not include the curvature quadratic terms with nontrivial $v_I$.} and defined ``pseudo-instantons'' as {\it metric-compatible gravitational field configurations with an irreducible curvature}, which solve the vacuum MAG field equations. The former condition means that the nonmetricity is trivial $Q_{\alpha\beta} = 0$ which for our wave ansatz means that $u = 0$ and $W^A = -\,V^A$. In terms of potentials this is translated into
\begin{equation}
{\mathcal U} = \overline{\mathcal U} = 0,\qquad {\mathcal W} = -\,{\mathcal V},\qquad
\overline{\mathcal W} = -\,\overline{\mathcal V},\label{noQ1}
\end{equation}
or equivalently
\begin{eqnarray}
{\mathcal X}_0 = 2{\mathcal W},\quad {\mathcal X}_1 = U - {\mathcal W},\quad
{\mathcal X}_2 = {\mathcal X}_3 = 0,\\ 
\overline{\mathcal X}{}_1 = -\,\overline{\mathcal W},\quad
\overline{\mathcal X}{}_2 = \overline{\mathcal X}{}_3 = 0.\label{noQ2}
\end{eqnarray}
In the pure quadratic model, $a_0 = 0$, whereas $w_I$, $z_J$, and $v_K$ are nonvanishing. Under the conditions (\ref{noQ1}), the field equations (\ref{eq1})-(\ref{eq8}) then reduce to
\begin{equation}
\Delta{\mathcal W} = 0,\qquad \Delta\overline{\mathcal W} = 0.\label{DWW}
\end{equation}
By making use of (\ref{pW}), we thus find $\partial_A W^A = 0$ and $\eta^{AB}\partial_A\underline{W}_B = 0$, and consequently from (\ref{OMAB1})-(\ref{OMAB4}) and (\ref{noQ1}) we conclude that 
\begin{equation}\label{nOm}
{}^{(2)}\!\!\mOm{\!}^\alpha = 0,\quad {}^{(4)}\!\!\mOm{\!}^\alpha = 0,\quad {}^{(I)}\!\!\pOm{\!}^\alpha = 0.
\end{equation}
In other words, the curvature is indeed irreducible (i.e., represented by just one irreducible part)
\begin{equation}
R_\alpha{}^\beta = {}^{(1)}\!W_\alpha{}^\beta,\label{Rir} 
\end{equation}
and we thus recover the pseudo-instanton solution in the sense of \cite{vas1}. It should be noted that the resulting gravitational waves are different from the ``torsion wave'' configurations described by Vassiliev.

\section{Discussion and conclusions}\label{DC}

In this paper we have studied the gravitational waves in the framework of the metric-affine theory for the class of models with the most general Lagrangian constructed from all possible parity-even quadratic invariants of the curvature, torsion and nonmetricity (\ref{LRT}). We have derived exact solutions with the help of the $pp$-wave type ansatz for the coframe (\ref{cof0})-(\ref{cof23}) and the linear connection (\ref{conW}). This ansatz gives rise to a very special structure for the curvature, torsion and nonmetricity which was clarified in Sec.~\ref{irreps}. 

We have solved the MAG field equations (\ref{dVt}) and (\ref{dVG}) in vacuum, i.e. under the assumption that the energy-momentum $\Sigma_\alpha = 0$ and the hypermomentum $\Delta^\alpha{}_\beta = 0$ matter currents are both vanishing.

It was shown that the plane-fronted wave solutions of GR and also of the teleparallel gravity arise as special cases of our solutions. For the latter theory, the resulting wave geometries are of the general type (with the vanishing curvature) and they encompass the two special subcases either with zero nonmetricity or with zero torsion (the standard teleparallel and the so-called symmetric teleparallel case, respectively).

The general gravitational wave solution in the class of quadratic MAG models is described by the fundamental transversal vectors $W^A$, $V^A$ and $u^A$, or equivalently, by the corresponding six scalar potentials ${\mathcal W}, {\mathcal V}, {\mathcal U}$ and $\overline{\mathcal W}, \overline{\mathcal V}, \overline{\mathcal U}$. Together with the function $U$ from the $pp$-wave metric ansatz, they constitute seven unknown functions which satisfy the system of eight equations (\ref{eq1})-(\ref{eq8}). Quite remarkably, the structure of the field equations demonstrates the complete decoupling of the parity-even and the parity-odd variables which satisfy the two separate sets of equations. In fact, this decoupling is explained by the absence of the parity-odd sector in the theory (no parity-odd coupling constants enter the Lagrangian (\ref{lagrV}) of the MAG model under consideration). Let us recall, in this relation, that there is no decoupling in the general quadratic Poincar\'e gauge gravity models with both the  parity-even and the parity-odd sectors included \cite{yno:2017,BCO}. Furthermore, one of the field equations, (\ref{2nd212}) (equivalently (\ref{eq3}) in terms of the potentials), fixes the dependence on the coordinate $\sigma$, whereas the remaining  system of seven Helmholtz or screened Laplace equations determines the seven unknown potentials as the functions of transversal coordinates $x^A$.

This system can be solved by making a standard exponential substitution (\ref{XX}) for the seven variables, which recasts the system into an algebraic form for the wave amplitudes. At this point, the parity-odd sector admits a more straightforward general analysis, whereas for the parity-even sector we have focused on certain particular physically interesting solutions to demonstrate the structure of the corresponding mode spectrum. Of special interest is the class of Yang-Mills type models in which the Lagrangian is constructed only from the curvature invariants, whereas the quadratic in torsion and nonmetricity terms are set to zero. Then the gravitational wave solutions represent the MAG ``pseudo-instantons'' in the sense of \cite{vas1}. It is worthwhile to stress that most of our results were obtained without or under very mild restrictions imposed on the parameters (coupling constants) of the action, so that the resulting geometries are exact solutions for large families of MAG models and not for specific sets of parameters.

Finally, it is important to remark that in our analysis we assumed the vanishing cosmological constant, the inclusion of which would require a serious modification of the plane wave ansatz for the coframe (along the lines of \cite{ndim} and \cite{BCO}), and the parity-odd sector in the gravitational action was not included here, despite the fact that possible violation of parity is widely discussed in the current literature \cite{Chen,Ho1,Ho2,Ho3,Ho4,Diakonov,Baekler1,Baekler2,Iosifidis,Obukhov:2021}. These issues remain open at the present stage of our research. We also have to postpone for the future the study of (possibly, simplified) models with a realistic matter distributions such as an in-falling dust or collapsing spheres of relativistic particles, e.g.

\begin{acknowledgments}
The work of AJC is supported by the Spanish Ministry of Economy and Competitiveness through the PhD contract FPU15/02864 and the project FIS2016-78198-P. YNO is grateful to Friedrich W.\ Hehl (Cologne) for the constant support and encouragement, deep questions and helpful comments. For YNO this work was partially supported by the Russian Foundation for Basic Research (Grant No. 18-02-40056-mega).
\end{acknowledgments}

\appendix

\section{Irreducible decomposition of the torsion}\label{irrtor}

The torsion 2-form can be decomposed into the three irreducible pieces, $T^{\alpha}={}^{(1)}T^{\alpha} + {}^{(2)}T^{\alpha} + {}^{(3)}T^{\alpha}$:
\begin{eqnarray}
{}^{(1)}T^{\alpha}&:=& T^{\alpha}-{}^{(2)}T^{\alpha} - {}^{(3)}T^{\alpha},\label{iT1}\\
{}^{(2)}T^{\alpha} &:=& {\frac 13}\vartheta^{\alpha}\wedge T,\label{iT2}\\
{}^{(3)}T^{\alpha}&:=& -\,{\frac 13}{}^*(\vartheta^{\alpha}\wedge\overline{T}),\label{iT3}
\end{eqnarray}
where the 1-forms of the torsion {\it trace} and {\it axial trace} are defined as
\begin{equation}\label{TT}
T := e_\nu\rfloor T^\nu,\qquad \overline{T} := {}^*(T^{\nu}\wedge\vartheta_{\nu}).
\end{equation}

\section{Irreducible decomposition of the nonmetricity}\label{irrnon}

The nonmetricity 1-form can be decomposed into the four 
irreducible pieces,
\begin{eqnarray}
{}^{(2)}Q_{\alpha\beta}&:=&{2\over 3}\,{}^*\!(\vartheta_{(\alpha}\wedge
\overline{\Lambda}{}_{\beta)}),\label{Q2}\\
{}^{(3)}Q_{\alpha\beta}&:=&{4\over 9}
\left(\vartheta_{(\alpha}e_{\beta)}
\rfloor\Lambda - {1\over 4}g_{\alpha\beta}\Lambda\right),\label{Q3}\\
{}^{(4)}Q_{\alpha\beta}&:=&g_{\alpha\beta}Q,\label{Q4}\\
{}^{(1)}Q_{\alpha\beta}&:=&Q_{\alpha\beta}-{}^{(2)}Q_{\alpha\beta}-
{}^{(3)}Q_{\alpha\beta}-{}^{(4)}Q_{\alpha\beta}.\label{Q1}
\end{eqnarray}
Here the Weyl covector 1-form is $Q:={1\over 4}g^{\alpha\beta}Q_{\alpha\beta}$, whereas $\aQ_{\alpha\beta}=Q_{\alpha\beta} - Qg_{\alpha\beta}$ is the traceless piece of the nonmetricity; and we denoted
\begin{eqnarray}\label{Lam}
\Lambda_\alpha &:=& e^{\beta}\rfloor\aQ_{\alpha\beta},\quad \Lambda:=\Lambda_\alpha\vartheta^{\alpha},\\
\overline{\Lambda}{}_\alpha &:=& {}^*\!\left(\aQ_{\alpha\beta}\wedge\vartheta^{\beta} - {\frac 13}
\vartheta_\alpha\wedge\Lambda\right).\label{oLam}
\end{eqnarray}
It seems worthwhile to notice that the 2-form $\overline{\Lambda}{}_\alpha$ which describes ${}^{(2)}Q_{\alpha\beta}$, has precisely the same symmetry properties as the 2-form ${}^{(1)}T^{\alpha}$.

\section{Irreducible decomposition of the curvature}\label{irrcur}

We start by splitting the general curvature 2-form as $R^{\alpha\beta}= W^{\alpha\beta} + Z^{\alpha\beta}$ into the skew-symmetric $W^{\alpha\beta} := R^{[\alpha\beta]}$ and symmetric $Z^{\alpha\beta} := R^{(\alpha\beta)}$ parts, and then we decompose the latter separately. The skew-symmetric piece is decomposed $W^{\alpha\beta} = \sum_{I=1}^6\,{}^{(I)}\!W^{\alpha\beta}$ into the 6 irreducible parts:
\begin{eqnarray}
{}^{(2)}\!W^{\alpha\beta} &:=& -\,{}^*(\vartheta^{[\alpha}\wedge
\overline{\Psi}{}^{\beta]}),\label{curv2}\\
{}^{(3)}\!W^{\alpha\beta} &:=& -\,{\frac 1{12}}\,{}^*(\overline{X}
\,\vartheta^\alpha\wedge\vartheta^\beta),\label{curv3}\\
{}^{(4)}\!W^{\alpha\beta} &:=& -\,\vartheta^{[\alpha}\wedge\Psi^{\beta]},\label{curv4}\\
{}^{(5)}\!W^{\alpha\beta} &:=& -\,{\frac 12}\vartheta^{[\alpha}\wedge e^{\beta]}
\rfloor(\vartheta^\gamma\wedge X_\gamma),\label{curv5}\\
{}^{(6)}\!W^{\alpha\beta} &:=& -\,{\frac 1{12}}\,X\,\vartheta^\alpha\wedge
\vartheta^\beta,\label{curv6}\\
{}^{(1)}\!W^{\alpha\beta} &:=& W^{\alpha\beta} -  
\sum\limits_{I=2}^6\,{}^{(I)}W^{\alpha\beta},\label{curv1}
\end{eqnarray}
where we denoted
\begin{eqnarray}
X^\alpha := e_\beta\rfloor W^{\alpha\beta},\qquad X := e_\alpha\rfloor X^\alpha,\label{WX1}\\
\overline{X}{}^\alpha := {}^*(W^{\beta\alpha}\wedge\vartheta_\beta),\qquad
\overline{X} := e_\alpha\rfloor \overline{X}{}^\alpha,\label{WX2}\\
\Psi_\alpha := X_\alpha - {\frac 14}\,\vartheta_\alpha\,X - {\frac 12}
\,e_\alpha\rfloor (\vartheta^\beta\wedge X_\beta),\label{Psia}\\
\overline{\Psi}{}_\alpha := \overline{X}{}_\alpha - {\frac 14}\,\vartheta_\alpha
\,\overline{X} - {\frac 12}\,e_\alpha\rfloor (\vartheta^\beta\wedge 
\overline{X}{}_\beta).\label{Phia}
\end{eqnarray}

For the symmetric curvature, at first we split it into the trace $Z = Z_\alpha{}^\alpha$ and the traceless piece $\aZ^{\alpha\beta} = Z^{\alpha\beta} - {\frac 14}Zg^{\alpha\beta}$. As a result, we obtain five irreducible parts
\begin{eqnarray}
{}^{(2)}\!Z^{\alpha\beta} &:=& -\,{\frac 12}{}^*(\vartheta^{(\alpha}\wedge\overline{\Phi}{}^{\beta)}),
\label{curZ2}\\
{}^{(3)}\!Z^{\alpha\beta} &:=& {\frac 13}\,\vartheta^{(\alpha}\wedge e^{\beta)}\rfloor
(\vartheta^\gamma\wedge Y_\gamma)\nonumber\\
&& -\,{\frac 16}\,g^{\alpha\beta}(\vartheta^\gamma\wedge Y_\gamma),\label{curZ3}\\
{}^{(4)}\!Z^{\alpha\beta} &:=& {\frac 12}\,\vartheta^{(\alpha}\wedge\Phi^{\beta)},\label{curZ4}\\
{}^{(5)}\!Z^{\alpha\beta} &:=& {\frac 1{4}}\,Z\,g^{\alpha\beta},\label{curZ5}\\
{}^{(1)}\!Z^{\alpha\beta} &:=& Z^{\alpha\beta} -  
\sum\limits_{I=2}^5\,{}^{(I)}Z^{\alpha\beta},\label{curZ1}
\end{eqnarray}
where 
\begin{eqnarray}
Y^\alpha := e_\beta\rfloor \aZ^{\alpha\beta},\quad Y := e_\alpha\rfloor Y^\alpha = 0,\label{ZY1}\\
\overline{Y}{}^\alpha := {}^*(\aZ^{\beta\alpha}\wedge\vartheta_\beta),\quad \overline{Y} := e_\alpha\rfloor \overline{Y}{}^\alpha = 0,\label{ZY2}
\end{eqnarray}
and 
\begin{eqnarray}
\Phi_\alpha &:=& Y_\alpha - {\frac 12}\,e_\alpha\rfloor (\vartheta^\beta\wedge Y_\beta),\label{ZPsia}\\
\overline{\Phi}{}_\alpha &:=& \overline{Y}{}_\alpha 
- {\frac 12}\,e_\alpha\rfloor (\vartheta^\beta\wedge \overline{Y}{}_\beta)\label{ZPhia}.
\end{eqnarray}

An important notice: the notations used here are different from \cite{MAG} in that ${}^{(4)}\!Z^{\alpha\beta}$ and ${}^{(5)}\!Z^{\alpha\beta}$ are exchanged, however, the components of irreducible parts match in a consistent way.

\end{document}